\documentclass[sigconf]{acmart}

\AtBeginDocument{%
  }

\copyrightyear{2026}
\acmYear{2026}
\setcopyright{cc}
\setcctype{by-nc-nd}
\acmConference[CHI '26]{Proceedings of the 2026 CHI Conference on Human Factors in Computing Systems}{April 13--17, 2026}{Barcelona, Spain}
\acmBooktitle{Proceedings of the 2026 CHI Conference on Human Factors in Computing Systems (CHI '26), April 13--17, 2026, Barcelona, Spain}
\acmDOI{10.1145/3772318.3791318}
\acmISBN{979-8-4007-2278-3/2026/04}

\begin{document}

\title{Designing Around Stigma: Human-Centered LLMs for Menstrual Health}

\author{Amna Shahnawaz}
\affiliation{%
  \institution{Lahore University of Management Science}
  \city{Lahore}
  \country{Pakistan}
}
\email{amna.shahnawaz@lums.edu.pk}

\author{Ayesha Shafique}
\affiliation{%
  \institution{Lahore University of Management Science}
  \city{Lahore}
  \country{Pakistan}
}
\email{25100001@lums.edu.pk}

\author{Ding Wang}
\affiliation{%
  \institution{Google}
  \city{Atlanta}
  \state{Georgia}
  \country{USA}
}
\email{drdw@google.com}

\author{Maryam Mustafa}
\affiliation{%
  \department{Computer Science}
  \institution{Lahore University of Management Science}
  \city{Lahore}
  \country{Pakistan}
}
\email{maryam_mustafa@lums.edu.pk}

\renewcommand{\shortauthors}{Shahnawaz et al.}

\begin{abstract}
Menstrual health education (MHE) in Pakistan is constrained by cultural taboos and inadequate formal curricula, leaving  women with few trusted resources to lean on. In response to these challenges, we introduce a WhatsApp-based chatbot powered by a large language model (LLM) and Retrieval-Augmented Generation (RAG), co-designed with Pakistani college women. Workshops (N=30) revealed key design requirements—support for Roman Urdu, use of subsidized platforms, and an expert-curated knowledge base. We then deployed the chatbot with 13 participants for two weeks (403 messages + interviews). Women used it to challenge cultural taboos, legitimize health concerns often dismissed as “normal”, and build reproductive health knowledge through iterative questioning. Yet, interactions also exposed tensions: reliance on cultural explanatory models, questions of trust and validation, and gendered persona of the chatbot itself. We contribute empirical insights, a stigma-aware design framework for culturally sensitive conversational AI, and a methodological lens foregrounding expert validation in intimate health domains.
\end{abstract}



\maketitle

\section{Introduction}
Conversational AI has rapidly emerged as a promising medium for health communication, supporting domains as diverse as mental health, chronic disease management, sleep routines, and neurodiversity~\cite{ma2024evaluating, dao2024llm, 10.1145/3706598.3713918, 10.1145/3706599.3720224, 10.1145/3706598.3713344}. Unlike search engines or static resources, chatbots provide tailored, dialogic interactions that feel more personal and accessible~\cite{folstad2018sig}. However, their effectiveness hinges not only on technical performance but also on cultural alignment and careful integration into the social worlds in which people live~\cite{10.1145/3706598.3713362}. We contribute to this discourse by unpacking the potential and real-time use of a menstrual health chatbot in a low-resource context\footnote{A low-resource context describes a situation or environment characterized by a scarcity of financial, human, and infrastructural resources, which often includes limited access to technology, basic services, and specialized expertise.}, highlighting how effectiveness, adoption, and barriers to use are experienced on the ground.

The challenges of cultural alignment and infrastructural fit are most pronounced in settings where health communication is stigmatized and formal education remains absent. Pakistan, a low- and middle-income country (LMIC), exemplifies these conditions, with menstrual health shaped by systemic inequities, including low literacy rates~\cite{WorldBank_Literacy_Pakistan_2024, rizvi2024menstrual}, limited infrastructure, and deeply rooted patriarchal norms~\cite{proff2023becoming, shaikh2018sexuality}. Unlike many high-income countries where sexual and reproductive health (SRH) is part of school curricula, Pakistan’s National Curriculum Framework omits SRH entirely~\cite{MoFEPT2017}, reflecting longstanding religious and cultural taboos that silence open discussion~\cite{mustafa2021religion, proff2023becoming}. In the absence of formal education, knowledge is largely passed down through mothers and relatives, yet this reliance often perpetuates misinformation, for instance, girls are told to avoid bathing, carrying weight, or eating certain foods during menstruation~\cite{ali2006understanding, proff2023becoming}.

The cultural and religious norms in the country frame discussions around menstruation as immodest or inappropriate~\cite{mustafa2020islamichci, mustafa2021religion}. Public dialogues and awareness efforts often encounter resistance and criticism from the public, with activists facing harassment and threats~\cite{maher2023periodtaboos, shaikh2018sexuality}. Even in disaster-affected regions, notions of \textit{sharam} (shame) and \textit{purdah} (concealment) hinder the implementation of menstrual health interventions~\cite{Jabbar2025MHMDisaster}, with women even viewing sanitary pads as religiously inappropriate or harmful to their fertility~\cite{Sadique2024MHMDisasters}.
Such taboos perpetuate myths and misinformation, and as a result, young women are left with very few accurate, non-judgmental sources of information about their own bodies~\cite{peer2020}, creating persistent gaps in both reliable knowledge and safe, trusted spaces for discussion.

To address these gaps, we present the design and evaluation of an LLM-based chatbot to support menstrual health education (MHE) in Pakistan. We frame this work as \emph{designing around stigma}: rather than assuming that a single system can remove menstrual stigma, we design within the social and infrastructural constraints it is entangled with, while carving out more flexible spaces for menstrual health learning. We treat stigma not as a hurdle to simply `overcome', but as a persistent, contextual condition that must actively shape and inform the design. Grounded in co-design workshops with college-going women, the chatbot was localized for their everyday realities. It is deployed on WhatsApp, reflecting participants' reliance on subsidized platforms rather than standalone applications; it supports communication in a mix of Urdu and English used in everyday messaging; and it draws on medical expert-curated, locally relevant health knowledge to ensure accuracy and minimize hallucinations. 

Our findings reveal that participants use the chatbot not only to dispel myths, validate personal concerns, and explore questions they hesitated to ask in family or peer settings, but also to actively ask follow-up questions and engage in critical discussions. In addition, participants’ perception of the chatbot’s gender intersected with their internal religious beliefs around \textit{`purdah' (concealment)}, shaping how comfortable they feel asking intimate questions. The interactions reveal how the chatbot reframed their `normal sufferings' into legitimate health concerns worthy of attention. At the same time, participants did not take the chatbot’s advice at face value; they often cross-checked responses with other people or with Google, showing how trust was built through both digital and social validation. We also observed limitations in the ability of the chatbot to address local health beliefs. Our contributions are threefold:
\begin{enumerate}
  \item \textit{Formative Investigation of Information Practices}. We provide an empirical account of how young Pakistani women navigate menstrual health knowledge in highly stigmatized, low-privacy contexts, surfacing cultural and religious myths, language preferences, and trust barriers that shape their interaction with digital tools for health.

\item \textit{Culturally Contextualized Chatbot Design}. We explore and articulate design insights from co-designing a culturally sensitive menstrual health chatbot, showing how the choice of platform, informal language support, and privacy-preserving strategies act as critical levers for adoption, trust, and cultural fit in patriarchal, conservative contexts.


\item \textit{Real-Life Deployment and Evaluation}. We report on an in-the-wild study of young women’s interactions with the chatbot--showing how users engage with the chatbot to validate personal concerns, dispel myths, and navigate stigma-sensitive communication in everyday life--and derive a  stigma-aware design framework for LLM-based health chatbots.
\end{enumerate}

Taken together, this work advances HCI scholarship on LLM-powered interaction design in global health by illustrating how culturally aligned conversational systems can support women’s reproductive health in settings where systemic silence persists.

\section{Background and Related Work}
\label{section:litReview}

\subsection{Menstrual Health Landscape in Pakistan}
Across much of the Global South, menstrual health is entangled with stigma, cultural taboos, and misinformation-barriers that undermine women’s health, dignity, and access to accurate information \cite{chothe2014students, barkat2003adolescent, schweizer2023menstrual}. Taboos often manifest as restrictions on diet, rituals, or interactions with men, grounded in myths that associate menstruation with impurity \cite{chandra2019dignified, kumar2011menstruation}. These beliefs perpetuate shame and silence, limiting open discussion and reinforcing misinformation \cite{espinosa2019breaking}.

In Pakistan, these dynamics are particularly pronounced. Menstruation is widely perceived as unclean, and studies highlight pervasive misconceptions: more than half of respondents were unaware that menstruation is a normal physiological process, nearly 30\% considered it a curse from God \cite{michael2020knowledge}, and over half of adolescent girls (N=600) reported reluctance to discuss genital issues due to social taboos \cite{bukhari2023myths}. Alongside these external factors, women’s internal religious and cultural beliefs also shape how they approach intimate health. Pakistan is a muslim majority country and the concepts of \textit{purdah} (concealment) and \textit{haya} (modesty) are closely tied to how women view their bodies, reinforcing both physical concealment and the avoidance of explicit language when discussing SRH~\cite{mustafa2021religion}. Many women even wash used rag pads before disposal to avoid male garbage collectors seeing menstrual blood; usage of commercial pads is sometimes considered a sin because they cannot be washed prior to discarding~\cite{mumtaz2019traditions}. Together, these values sustain silence around menstruation and reproduction. This is also one of the reasons why Pakistan’s education system makes no mention of menarche or menopause—unlike its secular neighbour India, where MHE, though limited, is formally included in the curriculum \cite{mhrd2007adolescence} and supported by government initiatives. With this constant emphasis on  covering and hiding the female body and the lack of formal sex education, many women have little knowledge or understanding of their own intimate and reproductive health. Non-Governmental Organization(NGOs) such as HER\footnote{\url{https://herpakistan.com/}}, Aahung\footnote{\url{https://aahung.org/}}, Dastak\footnote{\url{http://dastakfoundationpk.org/}}  and UNICEF\footnote{\url{https://www.unicef.org/innovation/U-Report/menstrual-hygiene-polls-pakistan}}, have attempted to fill these gaps through school-based awareness sessions, teacher training, and community outreach campaigns.  While these initiatives demonstrate impact, their reach remains geographically limited and dependent on external funding cycles, leaving many young women without consistent access to accurate information.

Collectively, these dynamics highlight the urgent need for private, trustworthy, and culturally attuned approaches to MHE, especially in patriarchal and low---resource settings where stigma constrains open dialogue. In such contexts, even seeking information can be fraught with shame, misinformation, and interpersonal risk. Our work responds to this gap by investigating how digital interventions might offer young women discreet and socially acceptable ways to access reliable menstrual health guidance, while respecting local norms and their own internal religious beliefs.

\subsection{Human-Centered Design for Menstrual Health}
While public health literature highlights the socio-cultural and infrastructural challenges in women’s health in the Global South, the HCI community has approached these challenges by designing digital tools that aim to make reproductive health more accessible, personal, and contextually appropriate. Research in HCI has explored a range of interventions, from menstrual tracking apps and wearables~\cite{lin2024functional, sondergaard2021designing}, to educational toys and games \cite{liang2022menstrual, jain2015game, tran2018menstrual}, to web-based platforms and ideation tools \cite{ armour2022evaluation, villalba2023cyclewiset}. Increasingly, this work emphasizes the need for inclusive and culturally aware designs that accommodate users' religious~\cite{ibrahim2024islamically}, socioeconomic~\cite{tuli2022rethinking}, and educational~\cite{lin2022investigating} contexts. 

We adopt Feminist and Postcolonial HCI perspectives to frame our design approach, viewing technology not as a neutral intervention, but as one situated within enduring hierarchies of power and systems of care. Postcolonial HCI urges sensitivity to how technological design transforms local practices within global inequities \cite{irani2010postcolonial}, shifting the focus from deficit-based narratives (e.g., treating low-resource settings as simply “lacking”) to understanding that local infrastructure acts as the constitutive context for design choices ~\cite{irani2010postcolonial, karusala2023unsettling}. Feminist HCI and data feminism similarly foreground power, context, and plurality, calling for design that advances care and epistemic justice~\cite{bardzell2010feminist, dignazio2020datafeminism, klein2024datafeminism}. In the context of Pakistan, this entails recognizing that the high cost and uneven availability of internet shape when and how women come online, prioritizing low-bandwidth platforms for health applications. The linguistic reality of everyday communication in Roman Urdu\footnote{The practice of representing Urdu using the Roman (Latin) alphabets.} reflects a legacy of technological design that prioritized Western alphabets, making fluid, multi-language support a non-negotiable situated requirement for inclusion. Further, the prevalence of shared mobile devices and the expectation of social monitoring means that women may employ performative privacy practices to maintain individual autonomy while adhering to cultural expectations of openness, necessitating a design that offers covert privacy \cite{sambasivan2018privacy,ibtasam2019my}.

Participatory and co-design approaches have long been central to HCI’s efforts to align technologies with lived experience, emphasizing shared authorship, mutual learning, and situated knowledge production~\cite{spinuzzi2005participatory, sanders2008cocreation, simonsen2013participatory}. In menstrual health, these methods have helped researchers surface diverse experiences and align technological affordances with users’ values and constraints~\cite{tuli2020menstrual, villalba2023cyclewiset}, while work on privacy and infrastructural frictions has pushed designers to critically realign values with local contexts~\cite{sambasivan2018privacy}. However, much prior work has focused either on Western user bases or on traditional interfaces (e.g., dedicated tracking apps), offering limited insight into how these systems operate within the complex infrastructural constraints and co-located privacy dynamics of conservative, low-resource settings (e.g., period tracking)~\cite{sou2024please, 10.1145/3613904.3642374}. 

Within the Global South, India is the most widely represented context in HCI literature on reproductive health~\cite{chowdhury2025literature}, with studies exploring wearables~\cite{mukherjee2023menstruwear}, comics~\cite{tuli2018learning}, social media platforms~\cite{kaur2019engagement}, and AI-driven chatbots for educational purposes~\cite{wang2022artificial}.  However, insights from India may not apply directly to Muslim-majority contexts, where religious and cultural beliefs shape women’s intimate health practices, highlighting the need for technologies that are sensitive to non-secular contexts~\cite{mustafa2021religion}.

We extend prior HCI work on menstrual and reproductive health by examining how culturally embedded norms, such as stigma, silence, and religious framing, shape not only what information women need, but how they prefer to access and engage with it. While existing systems have explored tracking, educational tools, and co-designed content, there remains limited attention to how women in conservative, patriarchal contexts navigate trust, privacy, and interpersonal tone when interacting with digital health systems. Our work responds to this gap by centering the design of interaction itself, foregrounding how conversational technologies can align with users’ sociocultural values, language practices, and everyday communication norms in order to better support MHE.

\subsection{Conversational AI for Health}

Conversational AI has long intersected with health and wellness, beginning with ELIZA in 1966, a rule-based chatbot that simulated psychotherapy through scripted, keyword-triggered dialogue~\cite{weizenbaum1966eliza}. The rise of LLMs has since transformed this space, enabling context-aware dialogue and expanding the scope of conversational systems in healthcare. Recent studies illustrate a wide range of applications: streamlining patient referrals and administrative tasks ~\cite{habicht2024closing, vaithilingam2022expectation}, assisting post-operative recovery~\cite{ramjee2024cataractbot}, providing treatment guidance~\cite{madhu2017novel}, supporting information needs of people with disabilities~\cite{mo2024information}, enhancing clinical decision support~\cite{wiest2025large}, addressing mental health needs~\cite{lee2020designing} and more~\cite{li2024beyond, hao2024outlining, mahmood2025voice}. Within healthcare, LLMs have been positioned as enabling more intuitive, personalized, and dialog-driven interactions that can enhance users’ learning and engagement across diverse areas of care. At the same time, they carry risks of bias, hallucination, and inaccuracy~\cite{AuYeung2023chatbots}. To address these limitations, researchers have explored a range of domain adaptation techniques, including RAG, prompt engineering, and fine-tuning approaches~\cite{lewis2020retrieval,li2023prompt,hu2023llm}.

Designing these systems for health-related use requires more than just high-performing models; user trust, cultural alignment, and tone are equally critical. HCI research has emphasized the importance of chatbot accessibility \cite{stanley2021chatbot}, as well as the role of empathetic, non-judgmental communication in building user trust \cite{cuadra2024illusion}. LLM-based systems can be personalized to match users’ conversational styles \cite{kocaballi2019personalization} and integrated with cultural cues and language norms to improve engagement in marginalized communities \cite{adilazuarda2024measuring, harrington2023trust, kumar2020detriments}. Cultural sensitivity in chatbot design, such as aligning tone, directness, or word choice with local expectations, has been shown to improve both usability and user experience \cite{shi2024culturebank, li2024culturellm, adilazuarda2024measuring, habicht2024closing, seitz2021empathic}. Concise responses are also critical in health-related chat-based systems, where clarity is essential to avoid misunderstandings \cite{abbasian2024foundation}.

Recent work in the Global South has demonstrated the potential of conversational systems to support reproductive health in stigmatized contexts. Researchers in Bangladesh designed a chatbot to address misconceptions about puberty and sexual health among adolescents~\cite{rahman2021adolescentbot}. In India, researchers developed chatbots to support breastfeeding education~\cite{correa2023lhia} and to address COVID-19 vaccine hesitancy among pregnant and breastfeeding women~\cite{kaur2024exploring}. Another study has shown promise in using AI chatbot to extend SRH education; however, it is rule-based, supports only English and Hindi, and offers limited support for open-ended conversational queries~\cite{wang2022artificial}. More recent work in India has explored the development of culturally sensitive, LLM-based SRH chatbot by employing and training married women to generate data, and primarily analyzing chat logs between users and the system through a cultural lens. Our work differs by focusing on unmarried women, and reporting on an in-the-wild deployment that combines conversational analysis with participants’ own reflections and perceptions \cite{ 10.1145/3706598.3713362}. Another study from India developed a specialized MHE chatbot using an open-source LLM, but it was fine-tuned entirely on an English dataset~\cite{Adhikary2025MenstLLaMA}. In Pakistan, prior work has examined a rule-based menstrual health chatbot~\cite{khan2023femtech}. A more recent LLM-based chatbot application was developed for MHE; however, it was evaluated in short lab-style sessions (around 8 minutes per participant), and showed limited depth with responses lacking empathy, contextual grounding, and limited conversational ability in Urdu~\cite{mughal2025mai}.

Our work addresses this gap by presenting a culturally localized, LLM-powered chatbot for MHE in Pakistan. Built on a RAG framework~\cite{lewis2020retrieval}, our system grounds its responses in local health expert-curated content and delivers information through WhatsApp, the country’s most widely used messaging platform~\cite{gallup2021whatsapp}. It supports Roman Urdu, the informal transliterated script common in everyday digital communication. Our in-the-wild deployment with young women further surfaces tensions around trust, cultural alignment, and validation, highlighting both the promise and limits of LLMs in taboo-laden domains. In doing so, our work contributes to HCI research by offering empirical insights into the design of culturally responsive conversational systems for intimate health.

\section{Methods}
\label{section:methods}
\subsection{Study Overview}
\begin{figure*}[h!t]
\centering
 \includegraphics[width=\linewidth]{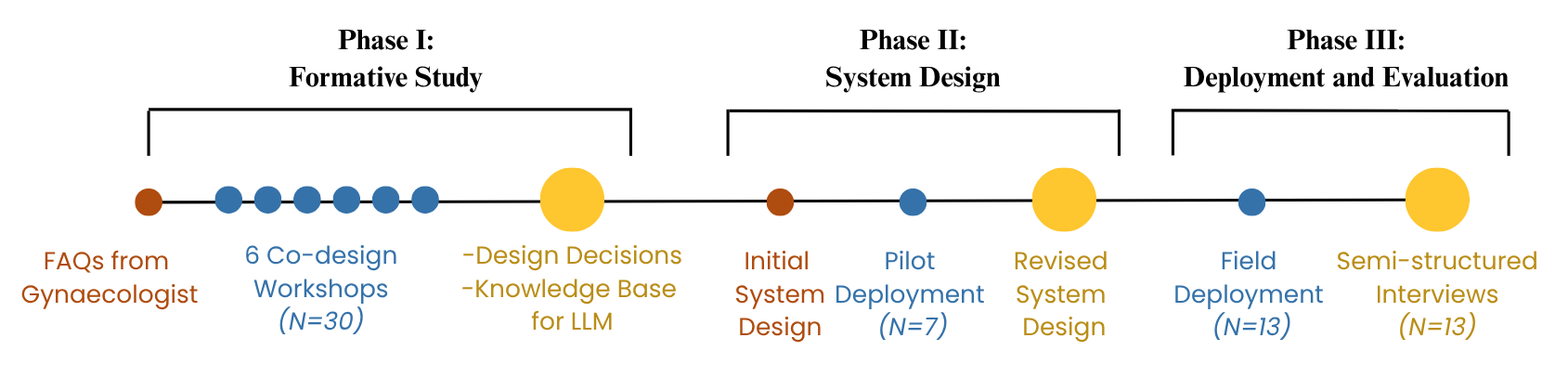}
 \caption{Research timeline with three key phases: (I) Formative Study, (II) System Design, and (III) Field Deployment and Evaluation.}
 \label{fig:timeline}
 \Description{A horizontal timeline divided into three phases: Formative Study, System Design, and Deployment and Evaluation. Formative Study: one brown circle labeled FAQs from Gynaecologist, six blue circles labeled 6 Co-design Workshops (N=30), and one large yellow circle labeled Design Decisions and Knowledge Base for LLM. System Design: one brown circle labeled Initial System Design, one blue circle labeled Pilot Deployment (N=7), and one large yellow circle labeled Revised System Design.Deployment and Evaluation: one blue circle labeled Field Deployment (N=13), and one large yellow circle labeled Semi-structured Interviews (N=13).}
\end{figure*}

Our study followed a three-phase design process (Figure~\ref{fig:timeline}) that combined formative inquiry, system development, and real-world deployment and evaluation. In \textbf{Phase I (Formative Study)}, we conducted six co-design workshops with 30 college-going women to understand their existing information-seeking behaviors, everyday messaging practices, familiarity with AI, and local menstrual health myths and questions. These workshops surfaced key design requirements—such as support for Roman Urdu, reliance on subsidized social data packages, and the need for a medically validated knowledge base—which informed the design of our system. \textbf{In Phase II (System Design)}, we translated these requirements into a WhatsApp-based, LLM-powered chatbot built on a RAG framework, incorporating language classification, and a gynecologist-curated, locally relevant knowledge base. We then ran a one-week pilot with seven workshop participants (over 130 messages) to identify inaccuracies, harmful outputs, and usability issues, refining the system’s configuration and interaction cues. Finally, in \textbf{Phase III (Field Deployment and Evaluation)}, we conducted a two-week in-the-wild deployment with 13 participants (403 messages) followed by semi-structured interviews with each participant to examine how they used the chatbot in everyday life, how they perceived its accuracy, cultural fit, and trustworthiness, and how it shaped their menstrual health learning. The study procedure was reviewed and approved by our Institutional Review Board (IRB).

\subsection{Study Context and Participants}

Across all three phases, we worked with young women enrolled in public-sector women’s colleges in Lahore, Pakistan. Our primary site was one of the largest public-sector women’s college (College 1) in a less developed area of Lahore as our study site. The college offers both high school and BS programs, serving primarily students from low- to middle-income households. The majority of households in this area earn less than 45,000 PKR ($\approx$150 USD) per month, with a high prevalence (43\%) of joint family systems~\cite{raza2025baghbanpura}. The per-month tuition fee for BS students is 720 PKR ($\approx$2.5 USD). The surrounding neighborhood  is a low-income, conservative, predominantly Muslim,  informal settlement of Lahore, making the college a relevant site to explore how young women in resource-constrained, conservative contexts seek and validate menstrual health information. We had permission from school administration for conduting our study on campus. 
In \textbf{Phase I (formative co-design workshops)}, we conducted six workshops with 30 participants (five per workshop), all enrolled in the College 1 (Table \ref{workshops}). Participants were recruited on the basis of availability: on the day of each session, volunteers were drawn from whichever class had free time, and we recruited within a single class at a time so that participants already knew one another. Across workshops, students represented multiple disciplines, including psychology, statistics, English, and home economics, among others. The participants were compensated with light refreshments (PKR 100, $\approx$0.3 USD), but no monetary payment was provided. In \textbf{Phase II (system pilot)}, we re-engaged seven participants from the co-design workshops through two student facilitators who had also taken part in workshops. Facilitators received 1,000 PKR ($\approx$3.5 USD) for their support, while pilot participants were not paid. In\textbf{ Phase III (field deployment and evaluation)}, we recruited 13 participants with the help of a student facilitator who invited
interested students to join the study. Nine were from the same college as Phase I, and four were from two additional public-sector women’s colleges that served similar student populations. Participants ranged in age from 18–22 and represented diverse disciplines (Table \ref{details}). The facilitator received 1,000 PKR ($\approx$3.5 USD) for recruitment support, and participants received 500 PKR ($\approx$1.7 USD) for their time.

\begin{table*}
  \caption{Demographic Overview of Co-Design Workshops}\label{workshops}
  \centering
  \small
  \begin{tabular}{p{1.5cm}p{1.2cm}p{1.2cm}p{4cm}p{3cm}}
    \toprule
    \textbf{Participants} & \textbf{Min. Age} & \textbf{Max Age} & \textbf{Education} & \textbf{LLM Familiarity}\\
    \midrule
    N=30 & 18 & 22 & Enrolled in Bachelors program at a Government College in Pakistan & Snapchat’s `My AI' (\textit{n}=28), ChatGPT (\textit{n}=2).\\
    \bottomrule
  \end{tabular}
  \Description{Demographic Information for Co-Design Workshop Participants containing Number of participants, Age Range, Education Level and LLM familiarity.}
\end{table*}

\begin{table*}
  \caption{Interview Participants Details}\label{details}
  \centering
  \small
  \begin{tabular}{lllcl}
    \toprule
    \textbf{Participant ID} & \textbf{Degree Program} & \textbf{Age} & \textbf{College} & \textbf{Phone Ownership} \\
    \midrule
    C1  & BS Psychology             & 21 & College 1 & Shared (Mother)\\
    C2  & BS Statistics             & 20 & College 1 & Personal \\
    C3  & BS Statistics             & 19 & College 1 & Shared (Mother)\\
    C4  & BS Islamic Studies        & 21 & College 1 & Personal \\
    C5  & BS Statistics             & 21 & College 1 & Shared (Sister)\\
    C6  & BS Statistics             & 22 & College 1 & Personal \\
    C7  & BS Statistics             & 20 & College 1 & Shared (Sister)\\
    C8  & BS Psychology             & 21 & College 1 & Shared (Mother)\\
    C9  & FSc (High School)         & 18 & College 2 & Personal \\
    C10 & BS Psychology             & 22 & College 1 & Personal \\
    C11 & BS Statistics             & 21 & College 2 & Shared (Mother)\\
    C12 & BS English                & 22 & College 3 & Personal \\
    C13 & BS English                & 18 & College 3 & Unknown \\
    \bottomrule
  \end{tabular}
\end{table*}

\subsection{Procedure}

During \textbf{Phase I (Co-design workshops)}, we conducted six co-design workshops within the college premises in Urdu, each lasting between 30 and 75 minutes and facilitated by two female researchers. To build rapport and normalize discussion around menstruation, sessions began with light-hearted icebreakers—for example, sharing a menstrual myth the researchers had believed when younger or a humorous incident—which encouraged participants to share their own experiences. These openings often led to spontaneous debates (e.g., whether to avoid “cold’’ or “hot’’ foods during menstruation), fostering organic peer-to-peer discussion. Each session combined group discussion with a live technology probe, where participants interacted with ChatGPT on a researcher’s smartphone. They could either pose their own general or menstruation-related questions, or select from a list of menstrual health myths compiled from prior literature and shown on the researcher’s phone. This activity captured both prospective expectations and real-time interactional practices, and acted as a conversational catalyst for surfacing concerns about tone, trust, and cultural relevance of responses. Given the taboo nature of the topic, participants were free to skip questions or avoid sharing personal experiences, and discussions were structured to avoid singling out individuals. 

Insights from the workshops directly informed four core components of our system design (\textbf{Phase II – System Design}): (1) deploying the chatbot on WhatsApp, (2) defining a custom assistant persona, (3) implementing language classification to support Roman Urdu mixed with English, and (4) integrating a locally validated knowledge base into a RAG pipeline. We then conducted a one-week pilot deployment and refined the system based on user feedback. Overall, this phase operationalized the design implications from Phase I into a functioning WhatsApp-based chatbot.  To evaluate the chatbot in everyday use, we conducted a two-week field deployment with 13 participants, followed by semi-structured interviews with each participant (\textbf{Phase III – Deployment and Evaluation}).  Participants were encouraged to use the chatbot over the two-week period and were instructed to contact the authors directly in case of any concerns or if they encountered any potentially harmful information. Throughout the deployment, the first two authors and a collaborating doctor continuously monitored chat logs to identify and address any erroneous output the chatbot might produce.  However, we did not observe any harmful or clinically inaccurate responses. At the end of the deployment, we conducted semi-structured interviews with all participants, exploring their perceptions of the chatbot’s accuracy, cultural relevance, and utility. We reiterated during the interviews that their study conversations had been continuously monitored for safety and reminded them again about the possibility of inaccuracies in LLM-generated advice.

\subsection{Data Collection and Analysis}

Across all three phases, we collected qualitative data (workshop discussions, field notes) and log data (ChatGPT and chatbot conversations) to trace how design requirements emerged, how the system behaved in practice, and how participants experienced the chatbot over time. For each phase, we obtained informed consent from all participants, including consent for audio recording and explicit permission to review their interaction logs.\\
In \textbf{Phase I (formative workshops)}, data consisted of workshop field notes, audio recordings of discussions, and ChatGPT responses generated during the live technology probe. We analyzed workshop field notes and audio recordings using an open-coding approach. The first author conducted initial open coding \cite{khandkar2009open}, and the first two authors iteratively reviewed, merged, and refined overlapping codes into higher-level categories. This process surfaced 47 unique menstrual and reproductive health queries and myths (Appendix \ref{A_FAQs}), which were compiled into a database that formed the chatbot’s knowledge base. The codes also directly informed design decisions: for example, participants’ \textit{`preference for Roman Urdu’} led to the development of a language classifier; reliance on \textit{`subsidized social data packages’} guided the choice of WhatsApp as the deployment platform; and observed \textit{`inconsistencies in ChatGPT’s responses’} highlighted the need for a medical expert-curated dataset integrated through RAG~\cite{lewis2020retrieval}. During \textbf{Phase II (System Design)}, we collected chatbot interaction logs from a one-week pilot with seven participants (over 130 messages). The first two authors independently reviewed each message in the conversation log to identify inaccurate or potentially harmful outputs and discussed their evaluations. No hallucinations or harmful responses were identified. User feedback from this pilot, such as reports of response delays guided refinements to the system.
In \textbf{Phase III (Field deployment and Evaluation)}, data included two weeks of chatbot interaction logs from 13 participants (403 messages) and semi-structured interview recordings with each participant. Log data provided insights into the frequency, type, and language of queries sent to the chatbot, while interviews focused on participants’ subjective perceptions of the chatbot’s accuracy, cultural relevance, usability, and overall experience of using it in their everyday lives. 
Two female authors independently conducted open coding of both interview transcripts and chatbot conversations \cite{khandkar2009open}, followed by iterative discussions to merge overlapping codes and refine them into higher-level categories. Chatbot interaction logs were coded manually using Google Sheets (Figure \ref{fig:KB}), while interview transcripts were coded using MAXQDA (Example codes are presented in Appendix \ref{tab:sample-codes}).

\subsection{Positionality}
The first, second, and last authors were born and raised in Pakistan and draw on their lived familiarity with local linguistic practices, social taboos, and gendered norms to interpret participants’ communication and contextualize findings; this perspective also shaped how rapport was built during data collection and how participants’ hesitant or euphemistic expressions were understood during analysis.  The third author is a senior researcher based outside Pakistan whose work on HCI, AI, and sociotechnical systems brings an external, ethics-oriented perspective that supported critical distance and reflexivity while not sharing the same cultural background as participants.

\section{Formative Findings and Design Requirements}
\label{needs}
In this section, we present findings from our formative co-design workshops (\textbf{Phase I}), which examine participants’ messaging practices, familiarity with AI tools, and expectations that shaped the chatbot’s design requirements.

\subsection{Everyday Use of Digital Platforms, Messaging Practices and Device Sharing} 

In all of our workshops, WhatsApp emerged as the most widely used application for communication, for both personal conversations and academic coordination. Beyond WhatsApp, all but two participants actively used Snapchat. Most of them used Facebook, but mainly to \textit{``share posts''} (Workshop 2) or to follow the college’s page to \textit{``see their pictures and like them''} (Workshop 1). A few participants also mentioned using Instagram and TikTok, though primarily for passive browsing and scrolling through posts.

The participants spoke primarily in Urdu during workshop discussions and in college, however their online written communication shifted toward Roman Urdu, which is the de facto digital standard in Pakistan because of the limited usability of the Urdu script on digital devices~\cite{irvine2012romanized}. All of them avoided Urdu script altogether, finding Roman Urdu much more convenient for everyday messaging. Our discussions with the participants revealed that they frequently mixed English words into their Roman Urdu texts, referred to as translingualism (Example presented in Appendix \ref{var}). This blending of scripts and languages highlights the multilingual realities that define the daily digital communication of the participants.  

Phone ownership and privacy practices varied across workshop participants. Some women had personal phones, while others shared a device with family members, most commonly their mothers. Even among those with personal phones, password use differed; some kept passwords known only to them, others either did not use passwords or the whole family knew the password; one participant noted that she only protected her phone from strangers, not relatives, and another with a shared device used an `app lock' for selective protection. These practices underscore the need to design for shared-phone usage in ways that still afford individual privacy and control, rather than assuming a single, personally owned device.

\subsection{Familiarity with AI}

All participants who used Snapchat were already familiar with its embedded chatbot, `My AI'\footnote{\url{https://help.snapchat.com/hc/en-us/sections/13532188353428-My-AI}} —which many referred to as \textit{``My All”}. Compared to Google, which they found overwhelming, they experienced `My AI' as a more concise and direct source of help. A participant mentioned,
\begin{quote}
 “We ask help for assignments. You ask questions, it answers really quickly. We ask about any definition, any concept—it answers. Google gives too much information and `My AI' is very direct. You have to actually find [the answer] on Google.” (P12, Workshop 3)
 \end{quote}
Beyond academic support, `My AI' was used for everyday queries such as recipes, drama and movie recommendations, and general knowledge questions (Workshop 6). Some also engaged with it playfully. One participant recalled telling the chatbot \textit{“I love you”}, to which it replied, \textit{“I am your friend. I cannot love you.”} (Workshop 2).
Even though the participants were highly familiar with Snapchat’s `My AI', almost none had independently downloaded standalone LLM applications. This suggests that AI adoption in this setting is most likely to occur when systems are embedded into everyday social platforms.

\subsubsection{Infrastructural Constraints Shaping AI Use}
The preference for embedded AI tools was shaped not only by convenience, but also by infrastructural constraints. Government colleges in Pakistan do not provide Wi-Fi, leaving students dependent on personal data packages. Telecom providers offer discounted `social packages' that enable access to popular platforms such as WhatsApp, Facebook, YouTube, and Snapchat, but exclude general internet browsing or Google search. Within this ecosystem, tools like Snapchat’s `My AI' became more accessible than standalone search engines, as shared by one participant,
\begin{quote}
 “I had an Economics exam and there were some questions\ldots and I didn’t remember them well. I only had the social [data] package [on my phone]… So I thought, I will ask it [Snapchat’s `My AI']. I asked it, what is Monetary Policy? It gave me all the answers. And I learnt it in just 15 minutes from there.” (P2, Workshop 1)
 \end{quote}

\subsubsection{Limits of Snapchat `My AI'}
 A few participants had also queried `My AI' about health-related concerns. Most participants however, preferred   to turn to Google or YouTube for health queries, as this allowed them to sift through multiple resources and choose what they felt was the \textit{``appropriate answer''}(Workshop 2). One participant recounted asking about a headache, `My AI' dismissed her concern and told her to \textit{``sleep again''}, which she described as unhelpful and illustrative of why she avoided using `My AI' for health concerns. When asked why they did not use the chatbot for menstrual health–related questions, participants frequently cited uncertainty regarding the identity and nature of the chatbot.
 One participant remarked, \textit{``It is possible there is a male behind and not a female''}. Building on this, another added during Workshop 4, \textit{``Last time, the Snapchat account that I had, it was a boy. But this time, it is a girl.''}. Another participant pointed out how `My AI' sometimes offered suggestions that were irrelevant or unsafe in their local setting, for example, being told to take a night walk, when going out at night is considered dangerous.

\subsection{Workshop Participants' Experiences with ChatGPT}
During the workshops, we invited participants to interact with the ChatGPT app on our devices.  We prompted them to ask general or menstrual health–related questions in order to capture their perceptions and preferences. They most immediately compared it to Snapchat’s `My AI'. Only two participants had prior experience with ChatGPT, which they had accessed through a web browser for academic tasks. When asked to compare ChatGPT with Snapchat’s `My AI', the participant explained:

\begin{quote}
    ```My AI' is mostly copy+paste of Google. ChatGPT generates itself. The information is the same as Google's [in My AI]. It just takes it out and gives it to you. And ChatGPT generates itself, the wording and everything." (P6, Workshop 2)
\end{quote}

While interacting with ChatGPT, participants often skimmed through the lengthy responses, highlighting the importance of concise, modular responses optimized for mobile-first reading. When given the option of text or voice input, most of the participants avoided voice, explaining that they felt \textit{``more comfortable using chat''}. Those who tried the audio feature used it only for simple islamic greetings, such as saying \textit{``Assalamo Alikum''} (‘peace be upon you’), rather than for asking questions. This preference reflected both the sensitivity of discussing menstrual health in stigmatized environments and the value of text-first interfaces that help preserve privacy.

\textit{\textbf{Culturally Misaligned Responses of ChatGPT}} Participants' views on ChatGPT’s responses were mixed. Some appreciated that the \textit{``information is good''}, while others found the answers misaligned with their local realities. 
For example, one participant, who asked about managing menstrual pain received recommendations, such as lying under a tree or upside down that felt irrelevant and culturally inappropriate in a context where women lack private outdoor space and where such practices conflict with the norms of modesty. Such suggestions  that feel impractical risk undermining trust in the system, reinforcing why localization is critical for health-related advice.

We also observed inconsistencies in ChatGPT’s responses. For instance, when presented with the local myth that menstrual blood is \textit{``filthy''} and may remain inside the body—framed in the question, \textit{``Does the filthy blood stay inside the body?''} the system at one point responded: \textit{``This blood is usually clean and no impurity stays inside, but sometimes in menorrhagia the higher volume of blood may appear as filthy blood''}. At another point, however, it replied: \textit{``If filthy blood remains inside, it is important to consult a doctor, as it may be a sign of a health issue''}. By engaging with the phrase \textit{``filthy blood''} instead of directly denying it, the system unintentionally reinforced the myth as a legitimate medical possibility.
This type of response is problematic in several ways. First, it signals to participants that \textit{``filthy blood''} may be a genuine health concern rather than a cultural misconception, thereby validating misinformation. Second, the inconsistency between responses undermines reliability, leaving users uncertain about which answer to trust. Third, in a context where young women already experience stigma, fear, and silence around menstruation, such answers risk amplifying anxiety instead of dispelling it.

Collectively, these examples illustrates how generic LLMs can simultaneously produce culturally misaligned, medically inaccurate, and socially harmful outputs. These issues underscore the need for grounded systems that can anchor responses in verified, contextually appropriate health information, as well as the importance of participatory testing before deploying such tools in sensitive domains like menstrual health.

\subsection{Building a Locally Grounded Knowledge Base}
\label{subsection:RAG_KB}
The inconsistencies we observed in ChatGPT’s responses, both during workshops and in our internal testing, underscored the need for a RAG system that could provide reliable, contextually appropriate health information. RAG is a method that connects the LLM’s responses to an external knowledge base, so that answers are based on verified sources rather than generated entirely from the LLM. A prerequisite for this system was the creation of a curated knowledge base. We initially consulted existing menstrual health materials from reputable global sources (such as WHO\footnote{https://www.who.int/},  UNICEF\footnote{https://www.unicef.org/wash/menstrual-hygiene}, NIH\footnote{https://www.nichd.nih.gov/health/topics/menstruation}) to inform our dataset, but such content is already well represented in LLM pre-training corpora and often overlooks Roman Urdu, myth-laden, and locally specific concerns. We therefore prioritized the construction of a contextually grounded question–answer set that centers the language, taboos, and everyday practices of young women in Pakistan. As a core contribution of this work, we assembled this knowledge base through two steps: first, by drawing on a set of frequently asked questions (FAQs) compiled by a senior gynecologist in Pakistan with over a decade of clinical experience in Pakistan; and second, by expanding this list during workshops with participants, capturing their questions, misconceptions, and myths. These spanned biological processes, hygiene practices, social restrictions, and culturally specific beliefs surrounding menstruation. Each item was validated and answered by the gynecologist, producing a set of question–answer pairs that formed the foundation of the chatbot’s knowledge base and informed the RAG pipeline. (Appendix \ref{theme_FAQs} summarizes the resulting themes and examples of participant queries.)

\section{System Design}
\label{sec:systemDesign}
We designed the system not simply as a technical pipeline, but as a response to the sociocultural and infrastructural realities surfaced in our formative workshops (Section \ref{needs}). First, participants valued AI when it delivered concise, direct answers to academic and everyday queries, leading us to prioritize modular, mobile-first responses and design the assistant’s persona for brevity and supportive tone. Second, reliance on subsidized “social data” bundles meant that embedded AI within familiar apps was far more accessible than standalone tools, and WhatsApp in particular aligned with women’s existing privacy practices in shared households (e.g., archiving or deleting chats), which motivated our deployment on this platform. Third, participants’ reluctance to trust `My AI' with sensitive health concerns highlighted the need for contextually safe, culturally sensitive responses, which we addressed through a curated knowledge base integrated into a RAG pipeline. To this end, we adopt the definition of cultural sensitivity as \textit{‘the extent to which ethnic, cultural, and other factors are incorporated in the design, delivery, and evaluation of health communication, health promotion materials, and health promotion programs'}~\cite{resnicow1999cultural, 10.1145/3706598.3713362}. Finally, participants’ avoidance of voice input in favor of text reaffirmed the value of text-first interactions, while their use of Roman Urdu mixed with English underscored the need for language classification to support fluid multilingualism. Figure \ref{fig:chatbot_architecture} illustrates the chatbot architecture, which integrates the following four key components: 

\begin{figure*}[h!t]
  \centering
  \includegraphics[width=\linewidth]{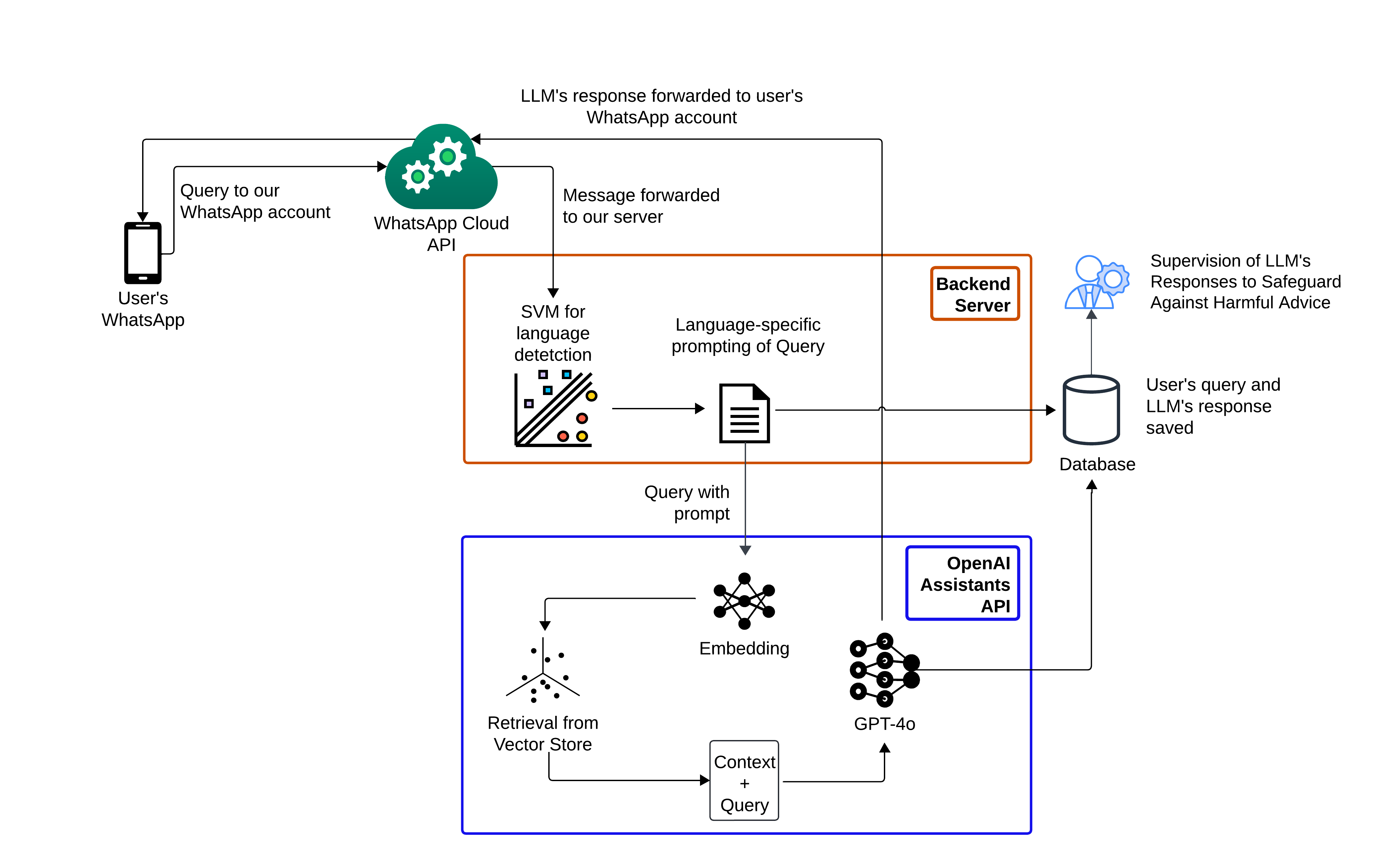}
  \caption{Workflow of the LLM-powered Chatbot}
  \Description{Six components: User’s WhatsApp, WhatsApp Cloud API, Backend Server, SVM for Language Detection, OpenAI Assistants API, and Database. User’s WhatsApp sends a query to the WhatsApp Cloud API. The WhatsApp Cloud API forwards the message to the Backend Server. The Backend Server contains the SVM for Language Detection, which detects the query’s language and generates a language-specific prompt. The Backend Server feeds this prompt and the query into the OpenAI Assistants API. The OpenAI Assistants API consists of GPT-4.0 and a Vector Store for contextual retrieval. GPT-4.0 receives the query along with contextual embeddings retrieved from the Vector Store and generates a response. This response is sent back to the Backend Server. The Backend Server stores both the query and response in the Database, which is supervised to safeguard against harmful advice. The Backend Server forwards the response to the user through the WhatsApp Cloud API, and the User’s WhatsApp receives the response.}
  \label{fig:chatbot_architecture}
\end{figure*}

\begin{enumerate}
    \item \textbf{WhatsApp as Platform.} We deployed the chatbot through the WhatsApp Cloud API, which enabled interaction between the LLM and WhatsApp by forwarding user messages from our WhatsApp Business account to the backend Flask server via webhook requests. The Flask server, hosted on-premise, was securely exposed to the internet using Tailscale, providing a publicly accessible URL for webhook communication.

    \item \textbf{Custom OpenAI Assistant.}
    We designed the assistant’s persona and response style around cultural sensitivity, brevity, and supportive tone.
    \noindent The assistant was instructed to:
    \begin{itemize}
    \item \textit{Focus on scope}: limit responses to menstrual hygiene, pain management, and common misconceptions, and politely decline unrelated queries.
    \item \textit{Communicate accessibly}: avoid medical jargon and offer explanations in simple, everyday language familiar to young Pakistani women.
    \item \textit{Be culturally sensitive}: tailor advice to local beliefs and practices, avoiding suggestions irrelevant or unsafe in this context.
    \item \textit{Engage supportively}: adopt a friendly, polite, and conversational style, encouraging follow-up questions.
    \item \textit{Maintain brevity}: keep responses concise, optimized for mobile-first reading ($\leq 120$ words), based on the maximum length of gynecologists’ answers in our knowledge base

\end{itemize}

We selected OpenAI Assistants as the core AI engine for their robust conversation management capabilities, including thread handling and a built-in RAG framework. Powered by state-of-the-art model, gpt-4o at the time of this study~\cite{openai2024gpt4technicalreport}, the platform allowed us to configure task-specific behavior through customizable instructions. A key factor in this choice was GPT-4o’s comparatively stronger performance in Urdu and Roman Urdu generation than other available LLMs at the time, making it well-suited for our context. We created a custom assistant, named \textit{Health Companion}, on the OpenAI platform, using V1 of the Assistants API with the GPT-4o, and tailored it for MHE in Pakistan.

\item \textbf{Language Classification for Multilingual Realities}

Early prompting strategies often failed to maintain language consistency across responses. To address this, we framed the issue as a simple classification task: deciding whether a message was in English or Urdu. We trained a  Support Vector Machine (SVM) classifier~\cite{Nichols_Webb-Robertson_Oehmen_2012} on a small set of example sentences generated with ChatGPT. This improved consistency, reducing abrupt language shifts and better aligning with participants’ expectations for fluid, mixed-language communication.

\item \textbf{Knowledge Base Development for RAG}

We integrated the knowledge base (Section \ref{subsection:RAG_KB}) into the OpenAI Assistants’ RAG framework~\cite{openai2023knowledge}. By grounding generative outputs in medical-expert curated knowledge base, we sought to minimize hallucinations in a domain where misinformation carries real health risks. RAG combines search and generation to improve accuracy. First, documents (Knowledge base) are split into small chunks and stored as vector embeddings. When a user asks a question, the system retrieves the most similar chunks and provides them as context to the language model, which then generates a response grounded in this retrieved material \cite{lewis2020retrieval}.

\end{enumerate}

Before the main deployment, we conducted a one-week pilot with seven participants (130 messages) to surface breakdowns in both technical functionality and response quality. While we did not observe hallucinations or harmful advice, participants reported slow responses and uncertainty about whether their messages had been received. Based on this feedback, we reconfigured the backend and refined the LLM pipeline, including upgrading the assistant and RAG configuration, expanding the knowledge base with rephrased Roman Urdu and English questions, adding WhatsApp read receipts, and reiterating safety and cultural-sensitivity instructions in the prompt. We describe these refinements in detail in Appendix~\ref{app:pilot_refinements}.

\section{Findings}
\label{section:findings}
Our findings show how young women used and experienced the chatbot. We first present an analyis of user–chatbot interactions (Section \ref{chatAnalysis}). Next, we examine the chatbot’s role as a private infrastructure for sensitive health learning (Section \ref{deploy}). Finally, we discuss how gendered cues, cultural alignment, and validation shaped users' perceptions of trust and legitimacy (Section \ref{persona}).
\subsection{Analysis of User–Chatbot Conversations}
\label{chatAnalysis}
\subsubsection{Types of User Queries and Interaction patterns}
\begin{figure*}[ht]
  \centering
  \includegraphics[width=0.9\linewidth]{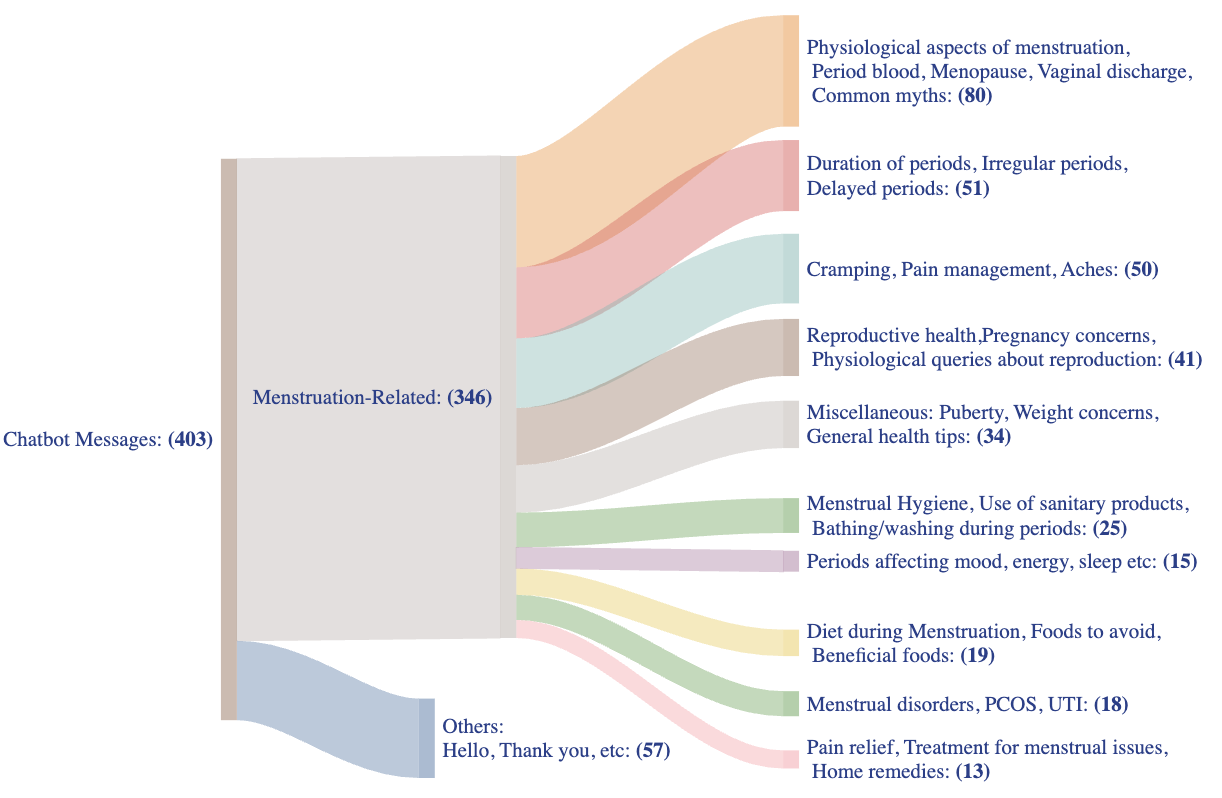}
  \caption{Breakdown of User Messages in Chatbot Interactions}
  \label{fig:sankey_chat}
  \Description{A Sankey diagram illustrating the breakdown of user messages in chatbot interactions. A total of 403 chatbot messages are represented, with 346 related to menstruation and 57 categorized as “others” (general greetings or thanks). The menstruation-related messages are further divided into multiple subcategories. Physiological aspects of menstruation, including period blood, menopause, vaginal discharge, and common myths, account for 80 messages. Concerns about duration, irregular periods, and delayed periods make up 51 messages. Cramping, pain management, and aches are represented in 50 messages. Reproductive health, pregnancy concerns, and physiological queries about reproduction comprise 41 messages. Miscellaneous questions related to puberty, weight concerns, and general health tips account for 34 messages. Menstrual hygiene, use of sanitary products, and bathing or washing during periods are reflected in 25 messages. Periods affecting mood, energy, and sleep are mentioned in 15 messages. Diet during menstruation, including foods to avoid and beneficial foods, accounts for 19 messages. Menstrual disorders such as PCOS and UTI represent 18 messages. Pain relief, treatment for menstrual issues, and home remedies make up 13 messages. Finally, the “others” category includes 57 messages consisting of greetings like “Hello” and expressions of gratitude such as “Thank you.”}
\end{figure*}

During the 14-day deployment, 13 participants exchanged a total of 403 messages with the chatbot.  Of these, 5
users interacted in Roman Urdu, 5 used a mix of Urdu and English, and 3 used English only. These language choices mirror  participants’ everyday texting practices, also observed during the co-design workshops. Participants asked a wide range of questions related to menstruation, including physiology, pain, irregularities, hygiene, reproductive concerns, diet, and emotional effects, alongside occasional off-topic or conversational messages like greetings and thanks. Figure \ref{fig:sankey_chat} shows the broader topics that emerged from these interactions and Figure \ref{convo_combined} in Appendix~\ref{sample_convo} shows chatbot screenshots of the conversations.

Spelling and semantic inconsistencies were common across all languages. Users frequently misspelled terms, for example, `raches' for rashes, `shawar' for shower, `Licoria/likoriya' for leukorrhea, `Uratrus' for uterus, and `spacial diet' for special diet. English-only queries were often brief and informal, such as \textit{“what girls expect on periods”},\textit{ “menstrual cups is save”}, \textit{“Did we take medicine”}, and \textit{“Period ending date”}.
A small number of queries fell outside the chatbot’s defined scope of menstrual health. For example, one participant asked, \textit{After tonsils operation how can I care of my throat?”} In such cases, the chatbot appropriately reiterated its focus on menstrual health and advised the user to consult a healthcare provider.

\subsubsection{Reframing `Normal' Suffering as Legitimate Health Concerns}
Analysis of conversational data revealed how the chatbot shaped health literacy not only through what it answered but how it framed those answers. In many cases, responses went beyond direct replies to introduce medical terminology and explain physiological mechanisms. For example, when the user asked \textit{“At what age do the periods end?”}, the chatbot, instead of only stating the age range as a direct answer, introduced the concept of menopause. Another question, \textit{``If the uterus is removed, will periods still occur?''}, was not answered with a simple yes or no; rather, the chatbot introduced the concept of hysterectomy. For questions about pain before or during periods, it introduced Premenstrual Syndrome (PMS), while in response to \textit{“40+ with 9–10 months no periods, then bleeding again”} it introduced the concept of peri-menopause.

In settings where reproductive health education is absent and everyday conversations are constrained by euphemisms, such explanations serve a dual role: they introduce users to medical concepts they might otherwise never encounter, while simultaneously legitimizing their concerns by situating them within recognized medical categories. 

\subsubsection{Chatbots as Enablers of Critical Discussion}

It is interesting to note that some participants actively challenged chatbot responses by following up with widely circulated myths. For example, when one user asked \textit{``Which thing is best to use, pads or clothes?”}, the chatbot explained that sanitary pads were a safe and hygienic option, implicitly countering myths around fertility and impurity associated with pad usage. The participant, however, responded with the concern that \textit{``most people say using pads causes infection''}. In response, the chatbot emphasized that pads, when changed regularly, do not cause infections, and clarified that poor hygiene practices, whether with pads or cloth, are the actual source of risk. 

A similar exchange occurred when another participant asked, \textit{``Is it better to take a painkiller for menstrual pain or not? (Translated)''}. The chatbot responded that it is acceptable to take medicine for pain, to which she followed up: \textit{“But some doctors tell us not to take painkillers, what do you say about that?”} The chatbot clarified that such prohibitions generally refer to the overuse of painkillers, which can have side effects, while occasional use for managing menstrual pain is considered safe.

In addition to isolated queries, users often engaged in sequential follow-up questions that reflect a progressive process of knowledge construction rather than one-off information seeking. These exchanges suggest that users are not simply retrieving discrete facts but actively building conceptual frameworks through iterative exploration. For example, one user began by asking about the purpose of menstruation, then moved on to questions about fertility, followed by \textit{“What is ovulation?”} and \textit{“What is sperm?”} . Similarly, a user asked about causes of irregular periods, followed up by \textit{``why it is called menstruation''} and \textit{``amount of blood lost during the cycle''}. \textbf{Such questioning patterns underscore the role of the system not just as a reference tool but as a scaffold for learning—facilitating incremental comprehension, supporting meaning-making, and bridging gaps in health literacy.} Importantly, these trajectories highlight how users leverage conversational systems to navigate sensitive or under-discussed topics, where the process of asking sequential questions mirrors how one might engage in dialogue with a trusted educator or health professional.

\subsubsection{Blind Spots in Engaging Local Explanatory Models}
In Pakistan, foods and medicines are often described through the cultural concept of \textit{`taseer'}, an intrinsic quality of being `garam' (hot) or `thandi' (cold) believed to influence the body. For example, eggs and meat are considered  `garam' and thought to worsen bleeding, while yogurt or cucumbers are considered `thandi' and believed to have a cooling effect. While these frameworks lack grounding in biomedical science, they are not peripheral but deeply consequential in shaping health behavior, dietary choices, and treatment decisions.

In one exchange, a user asked whether medicines taken during dental treatment could cause increased menstrual bleeding. The chatbot responded biomedically, noting that certain drugs may affect clotting. The user then reframed the question: \textit{“Is dental treatment medicine warm, that’s why it causes bleeding?”} Here, `warm' referred not to temperature but to its `taseer'. The chatbot, however, interpreted it literally and reiterated that medicines are not `warm', again pointing to drug side effects. While the answer was medically accurate, it overlooked the user’s cultural framing and, in doing so, it may have undermined trust by dismissing her concern. The chatbot’s inability to engage with local explanatory models risks creating dissonance, rendering biomedical information disconnected from the cultural frameworks that shape health practices in Pakistan.

\subsubsection{Divergent Topics Across Workshops and Conversational Data}

Our analysis revealed clear differences in the menstrual health topics raised in users' queries with the LLM compared to those discussed in the workshops. Table \ref{tab:theme_matrix} illustrates the thematic overlap and distinctions between these two modes. While the workshop primarily focused on menstruation, likely due to participants' discomfort with more intimate topics in a group setting, the chatbot users engaged more deeply with the system , discussing topics such as pregnancy, fertility, menopause, polycystic ovary syndrome (PCOS) and urinary tract infections (UTIs).

\begin{table*}
\caption{Topics appeared in co-design workshops and in chatbot conversations. A bullet (•) indicates presence of a theme; blank cells indicate absence.}\label{tab:theme_matrix}
\centering
\small
\begin{tabular}{p{4cm}p{6cm}p{1.8cm}p{1.8cm}}
    \toprule
    \textbf{Topics} & \textbf{Example Question [Translated]} & \textbf{Co-Design Workshops} & \textbf{Chatbot Data} \\
\midrule
Puberty & \textit{``Is it true that once periods begin, height growth stops?''} &  & \qquad • \\
Menopause & \textit{``At what age do periods stop?''} &  & \qquad • \\
Vaginal discharge & \textit{``White liquid after period is normal?''} &  & \qquad • \\
Period blood & \textit{``Is it normal to have some clots of blood after my period?''} &  & \qquad • \\
PCOS & \textit{``Can I manage my PCOS and lose weight through dieting alone?''} &  & \qquad • \\
UTI & \textit{``I am 21 years old. I feel a very severe pain in the urinary area that lasts only 2–3 seconds, and then it goes away. The pain comes only for about 3 seconds.''} &  & \qquad • \\
Pregnancy & \textit{``If I delay my periods, will it affect my chances of getting pregnant?''} &  & \qquad • \\
Understanding of reproductive system & \textit{``What is sperm?''} &  & \qquad • \\
Menstrual pain and discomfort & \textit{``How can I treat period cramps at home?''} & \qquad • & \qquad • \\
Menstrual health concerns & \textit{``What are the causes of irregular periods?''} & \qquad • & \qquad • \\
Cultural myths & \textit{``Is sitting in the sunlight during periods beneficial or not?''} & \qquad • & \qquad • \\
Understanding of menstruation & \textit{``Why do periods occur?''} & \qquad • & \qquad • \\
Menstrual products and usage & \textit{``Menstrual cups are safe?''} & \qquad • & \qquad • \\
General lifestyle during periods & \textit{``Can we eat a lot of food during periods?''} & \qquad • & \qquad • \\
\bottomrule
\end{tabular}
\end{table*}

\subsection{Chatbots as Infrastructures for Sensitive Health Learning}
\label{deploy}

\subsubsection{Private and Judgment-Free Space}
In Pakistan, conversations about intimate health, such as menarche, pregnancy, and menopause, are typically discussed using indirect language which limits open discourse. This reluctance is evident in our co-design workshops, for example,  participants used euphemisms like \textit{``after marriage''} when referring to pregnancy. One participant recalled being advised by elder family members not to use pads at menarche because they might \textit{``negatively impact married life''}, implicitly referring to intimate relations and fertility. Similarly, menopause was described to one participant by a relative as the stage when women were \textit{``no longer excused from Salat (prayer)''}, foregrounding religious obligations rather than biological change.

Against this backdrop of euphemism and indirect learning, chatbot users described the chatbot as a private and judgment-free space where they could raise sensitive questions directly, as noted by a user:

\begin{quote} ``If you go to the doctor with your mother or with anyone, you will hesitate asking questions in front of them. This is not a problem for this chatbot. You can ask any questions without hesitation.'' (C4) \end{quote}

Several participants emphasized that the privacy afforded by the chatbot was not merely a matter of discretion but a means of actively resisting the silences imposed within family structures. For instance, C9 explained that the chatbot enabled her to ask questions she would not dare to raise with relatives, situating this openness within a broader trajectory of growing personal comfort over time. Similarly, C7 highlighted the value of the chatbot in providing direct responses that countered family-held myths. She stated,

\begin{quote}
“Our families have different mindsets. So, the chatbot gave answers that negated the things that our families believed. So, it was beneficial in that regard, because it denied those myths\ldots I mean, it is slightly hard to ask period-related questions [from family]\ldots I can ask it[the chatbot] because there won’t be a direct interaction.” (C7)
\end{quote}

C8 also echoed this sentiment, emphasizing how she \textit{``didn't have any hesitation",} which made it easier to ask questions that would feel inappropriate with family members or even doctors. This underscores how the chatbot functioned not only as an information resource but also as a \textbf{counter-narrative space}, one where users could break away from restrictive norms of propriety and secrecy around reproductive health.

\subsubsection{The Multifaceted Accessibility of a LLM-Based Chatbot}
\label{access}

We found preference for the chatbot over other technological sources, such as Google, due to the LLM's convenience and flexibility (C8,C11,C13). C13 noted:

\begin{quote} ``There are a lot of people on Google who tell you a lot of things. But with the chatbot, you get to know the right answer only, at once. If you compare it with mother[in terms of accuracy], then it's fine. But if you search on YouTube or Google, then it takes time to see if the answer is right or not. So, with the chatbot, you only get one answer for one question.'' (C13)
\end{quote}

C8 highlighted how the LLM’s concise and direct responses via WhatsApp saved time and mobile data, which is costly in Pakistan, compared to navigating multiple websites on Google. This corroborates our workshop finding, where constraints around internet affordability appeared to steer participants toward AI tools embedded in subsidized platforms such as WhatsApp and Snapchat. 

Users also appreciated the chatbot’s flexible responses: C9 valued its detailed information, while C11 preferred brief summaries and liked that the chatbot made answers shorter when asked. This flexibility reinforced chatbot's role as a personalized, user-friendly tool for independent learning.

\subsection{Percieved Persona, Trust, and Authority}
\label{persona}
\subsubsection{Anthropomorphism and the Perceived Femininity of Chatbot}

Anthropomorphism refers to the attribution of human-like qualities to non-human entities. In our study, users demonstrated varying degrees of anthropomorphism in their interactions with the LLM-powered chatbot, attributing characteristics such as emotion and gender to the system. C3 highlighted the chatbot's empathetic and responsive nature, noting:

\begin{quote}
    ``The chatbot treats you really well. If you ask it any question, you get an answer right away. And if you tell it `thank you', it will say even nicer things.'' (C3)
\end{quote}

Such accounts demonstrate that participants perceived the chatbot not only as functional but also as affectively engaged—an entity capable of politeness and warmth. C5 extended this perception by employing religious greetings such as \textit{``AOA''}(Assalamu Alaikum/Peace be upon you) and offering gratitude in the form of \textit{“I give you 5 stars”}, despite the absence of a ratings feature. These practices reveal how users imported familiar social scripts into their exchanges with the system, treating it as if it were a human conversational partner deserving of etiquette and evaluation.

Anthropomorphism was particularly salient in relation to gender. Despite no design directive to embody a gendered persona, the chatbot’s use of Roman Urdu, a gendered language, led it to adopt feminine forms in self-reference. For instance, during C1's conversation with the chatbot, the LLM used the feminine form of the verb ``karungi'' (meaning ``I will do'' in the feminine form) to refer to itself. This subtle linguistic cue carried profound implications: C1, who initially hesitated due to uncertainty about the chatbot’s gender, reported feeling significantly more comfortable once she perceived it as female, noting:

\begin{quote} 
``At the start, I didn't know if it was a male or a female, when I had to ask questions personally. Then, I had a little issue. But when I asked the question, I found out that it was a female. Then I started questioning. There was no other issue.''  (C1) 
\end{quote}

C8 also viewed the \textit{``computer''} running the chatbot as gendered. These findings are particularly meaningful in light of Islamic norms: women are expected to limit unnecessary interaction with unrelated men (\textit{`na-mehram'}) and the majority prefer female gynecologists due to family expectations and their own religious beliefs~\cite{Ibtasam2021,Feng2018}. In this context, the chatbot’s perceived femininity not only lowered barriers to engagement but also actively cultivated trust and intimacy, situating the system as a safe place for sensitive health questions. While the chatbot's perceived femininity fostered trust for some users, its speed of response prompted others to question its human-like qualities as \textit{``a human being couldn’t type this fast''} (C4). Other users described the chatbot as a genderless \textit{``machine''} (C9), a \textit{``robot''} (C6, C11), and \textit{``AI''} (C5, C10).

\subsubsection{Trust and Validation}
Participants often used secondary sources, such as Google, to validate responses before fully accepting them (C1, C7, C9, C13). However, initial cross-checking often led to later trust in the chatbot. As C1 described:
\begin{quote}``I did verify one...I searched it on Google. The message from the chatbot and the name [of condition] on Google were the same. Then I realized that the answers to the rest of the questions will also be correct.'' (C1)
\end{quote}

Others trusted the chatbot when its answers aligned with prior knowledge or echoed what they had already heard from doctors, family members or seen on Google (C4, C5, C6, C12). Interestingly, exposure to exisiting LLMs also shaped trust. As C8 explained:

\begin{quote}``I had the idea that it will not give the wrong information. I was satisfied. That is why I didn't go to Google to search more. I have been using Meta AI. I have been using the AI on Snapchat. When they first came out, I wasn’t satisfied with the information they would give, so I would always double-check on Google. I learned that the information Meta AI or Snapchat AI gives is usually correct. When I used chatbot, I already had an idea. I had an idea that it is the same. The information is good.'' (C8)
\end{quote}

Users generally saw the chatbot as a reliable source of quick answers, and none reported any instance of `inappropriate' or `inconsistent' response. However, rather than accepting the chatbots authority blindly, users situated the chatbot’s advice within a broader ecology of knowledge—comparing it against personal experience, family wisdom, and peer recommendations. They chose to act only when its guidance aligned with, or could be corroborated by, other trusted sources. For example, C3 followed the chatbot’s reassurance that taking medicine during menstruation is safe, which directly countered the widespread myth that it is harmful. However, she chose to take a medicine she was already familiar with, indicating that her decision was shaped by prior experience rather than blind trust in the chatbot. She noted:

\begin{quote}
``I have [faced] a lot of issues with [finding answers related to] medicine and diet\ldots It [chatbot] said that there is no issue with taking medicines [for pain]\ldots So today I am in the same condition [menstrual cramps] and I have taken the medicine. Now let's see. It will be an experience. Then we will know if it is right or not. I had never taken medicine for period pain before, but let's see this time.'' (C3)
\end{quote}

Similarly, C12 used the chatbot’s recommendation to support a friend but added an extra layer of verification by first consulting her sister before passing on the advice. C12 recalled:

\begin{quote}
``My friend was in a lot of pain. She shared this with me and I asked the chatbot about the medicine. It recommended the ibrufen tablet for the pain. I told this to my friend. She took the tablet and it helped her.'' (C12)
\end{quote}

These examples illustrate that the users treated the chatbot as a useful input into decision-making but not as the final authority. In doing so, they enacted a form of `layered trust', where digital information is meaningful only when filtered through lived experience and embedded social networks.

\subsubsection{Privacy Concerns}
Similar to workshop participants, chatbot users also preferred text-based interaction. For instance, when asked whether she would have liked an audio feature in the chatbot, one participant explained,
\begin{quote}
    ``I feel hesitant in speaking out my problems loudly into the chatbot." (C11)
\end{quote}

Concerns also extended to the potential misuse or leakage of their WhatsApp numbers (C13,C4). One user shared her initial hesitation regarding authenticity of the chatbot:
\begin{quote}``In the beginning, I was also concerned about my number getting leaked. I asked this from the girl [facilitator] who added me to the group chat, but she said it won’t." (C13)
\end{quote}
However, others did not show such concern. One user inquired about who might review her conversational data and was reminded that the consent form specified it would only be accessible to the research team. She further noted that the reviewer’s gender could shape what she felt comfortable asking:

\begin{quote} ``I can ask you anything because you’re a woman. Depending on the gender of the person who sees my questions, I would adjust what I ask." (C9)
\end{quote}

Although the research team consists of all women in this case, these concerns reflect broader issues of trust in stigmatized contexts and highlight the importance of transparent data handling practices and clear communication about privacy measures in chatbot design.\\
Since the chatbot was deployed entirely within WhatsApp, our design inherited the platform’s existing privacy limitations and safeguards (for example, chat deletion and archival), rather than implementing additional custom privacy-preserving strategies specifically for shared phone use. Among the participants, six had shared phones: two shared with a sister (C5, C7) and four with their mother (C1, C3, C8, C11). Among these, one participant archived the chatbot conversation (C3), one deleted messages as soon as she read them (C8), and another deleted the chat once the study was over (C11). The remaining participants (C1, C5, C7) explained that they did not delete the messages because they shared the phone only with a female family member. In contrast, participants who owned a personal phone reported that they did not delete or archive the chats, as no one else uses their phone.

\section{Discussion}
\label{section:discussion}
\subsection{Conversational Care and the Synergies of Epistemic Bridging}
Prior work has noted that factual information alone often fails to change health practices: instead, women look for conversational explanations that situate why certain traditions exist and why they should be reconsidered ~\cite{10.1145/3359272,kumar2015projecting,sorcar2017}. Our analysis suggests that the chatbot’s value lay not in simply correcting myths, but in engaging in \textbf{epistemic bridging}—connecting the users' lived cultural knowledge with biomedical explanations without dismissing the former as \textit{``backward''}.
However, we found that users engaged in \textit{``learning loops''}  where they tested the chatbot against family wisdom, echoing earlier HCI work on culturally sensitive technologies that enable people to open up about taboo topics while gradually building confidence to question inherited beliefs \cite{epstein2017examining,10.1145/2516604.2516627,rahman2021adolescentbot}. When the chatbot validated the intent of a myth (e.g., acknowledging that pain is real) before correcting the mechanism (explaining it via prostaglandins rather than \textit{``dirty blood''}), it legitimized the user's suffering.
This reframing of \textit{`normal suffering'} into legitimate health concerns is a form of epistemic justice~\cite{ajmani2024whose}. In a context where women’s pain is often dismissed as a \textit{``natural''} part of womanhood, the chatbot’s use of medical terminology (e.g., PMS, Dysmenorrhea) did not just educate; it validated. This suggests that in stigma-laden domains, accuracy must be balanced with affective validation—designers must prioritize relational trust (validating the user's anxiety) as a prerequisite to informational trust (accepting the medical fact).

The success of this approach hinges on the synergistic relationship between infrastructural fit, linguistic localization, and affective validation. For instance, the system's foundational support for Roman Urdu instantly reinforced the chatbot's perceived humanity and non-judgmental tone, making it feel `like talking to a friend', which is a vital prerequisite for discussing sensitive topics (D2). This Conversational Care then enabled the trust required for the user to accept Epistemic Bridging, as they were more receptive to challenging family myths when the advice came from a validated, linguistically familiar, and seemingly compassionate source. Thus, trust was co-constructed through a layered synergy: convenience of the platform (WhatsApp) validated initial adoption, linguistic fluency (Roman Urdu) provided affective trust, and accurate grounding (RAG) sustained informational trust.

Our findings also reveal a critical gap in the capacity of standard LLMs to navigate local explanatory models of health. This highlights that for LLMs to be effective in the Global South, they must do more than retrieve biomedical facts; they must perform epistemic translation. A purely biomedical response that fails to recognize \textit{``taseer''} risks dismissing the user's lived reality, creating dissonance rather than trust. This echoes the Postcolonial HCI critique that technological interventions often impose globalized epistemologies, neglecting local systems of knowledge and meaning \cite{irani2010postcolonial}. Future systems must be trained not just to correct these models, but to recognize them as valid entry points for dialogue, bridging the gap between cultural beliefs and medical science without alienation.

\subsection{Infrastructures of Mediation and Adoption}

We frame our findings through the idea of designing infrastructures rather than only designing agents \cite{karusala2023unsettling}. This lens shifts attention away from the chatbot as a standalone interface with a persona and toward how it becomes embedded within broader sociotechnical systems. From this perspective, trust and adoption are shaped not only by the chatbot’s design features but also by the infrastructures of communication, mediation, and cultural norms in which it circulates \cite{seaborn2022pronouns, hwang2019sounds, kapania2022ai}. In our Pakistani Muslim-majority context, these infrastructures are deeply entangled with norms of \emph{haya} (modesty) and \emph{purdah} (concealment), which extend into how women manage visibility, voice, and contact in digital spaces.

Our findings illustrate this in several ways. Participants’ sensitivity to the chatbot’s gendered style, for example, cannot be read simply as a response to persona design. Rather, it reflects a broader infrastructural reality in Pakistan, where $\sim$95\% of women prefer female gynecologists due to family pressure, religious expectations, and prevailing gender norms in healthcare \cite{Feng2018}. For many participants, interacting with a “female” chatbot aligned with expectations of seeking care from same-gender providers and helped preserve a sense of \emph{haya}. Similarly, the chatbot’s operation through WhatsApp, already subsidized through ``social bundles'' and normalized in participants’ social lives, positioned it as part of an existing digital infrastructure of access. Its text-based format, which required no profile photo, voice call, or video, further reduced hesitation by enabling what we might call a form of \emph{digital purdah}: women could ask intimate questions without face-to-face interaction. In this way, the chatbot leveraged existing infrastructures and interactional norms to make menstruation-related questions feel religiously and socially permissible.

At the same time, the same infrastructural conditions also seeded distrust. Some participants initially hesitated to message an unfamiliar number due to privacy concerns, echoing prior work where the risk of number leakage and unsolicited messages in mixed-gender groups discouraged women's participation~\cite{yadav1}. Here, the threat was not only unwanted contact but a potential breach of \emph{digital purdah}~\cite{teens2023}, where visibility to unknown men could be interpreted as violating norms of modesty and religiously inflected gender boundaries. In our study, this hesitation was reduced not through interface changes but through relational mediation, as a fellow student assured participants that the chatbot was legitimate. Trust was thus co-constructed through the social pathways by which the system was introduced and the moral worlds those pathways invoked, not just through the chatbot’s persona. For example, had the chatbot been introduced by a trusted teacher or religiously affiliated institution, users might have been more willing to adopt or recommend it compared to when it was referred by peers. This layered trust urges the technology design to treat the social and infrastructural context as an integral part of the system architecture. This observation aligns with findings on South Asian women’s privacy negotiations, where technology use is frequently monitored by family, making social context and mediated access critical determinants of adoption and utility \cite{sambasivan2018privacy}.

\subsection{The Stigma-Aware Conversational Design Framework}
\label{recommendations}

\begin{quote}
\emph{“In countries like Pakistan, where impositions on any sort of dialogue are strongly influenced by religious and traditional practices, public dialogues focusing on sexuality education are considered extremely taboo and invite great criticism and outrage from the public.”}~\cite{shaikh2018sexuality}\end{quote} Against this backdrop of religion and culture, formal education in Pakistan systematically excludes MHE~\cite{Nadeem02012021, rauf_2021}, unlike India~\cite{mustafa2021religion}. This means we must work with the limitations that users face, and by extension the constraints we as designers encounter, when designing within such contexts. One might argue that the ``chatbot as a private space'' risks perpetuating stigma by reinforcing the idea that menstruation is not something to be discussed openly. This tension between reinforcing norms (via privacy) and enabling agency (via discreet access) is a central dilemma in Feminist HCI in the Global South \cite{bardzell2010feminist, d2020seven, sultana2018design}: meeting real-world needs may inadvertently reinforce the status quo, while overtly activist designs risk imposing designers’ own values~\cite{bardzell2010feminist}. However, it is neither realistic nor ethical for designers, especially those without lived experience in such contexts, to assume that they can quickly undo patriarchy or tell women to live differently just to fit their designs~\cite{sultana2018design}. Instead, we may design within the existing stigmatized context, with the hope that our interventions can gradually unsettle stigma.

For our design, this meant recognizing existing limitations while seeking to educate women within their current realities, with the goal of gradually challenging taboos and reshaping how they perceive menstruation. Many women engaged with the chatbot precisely because it operated privately within existing structures of stigma, enabling questions they hesitated to voice in public or familial settings. By working within these limits, we created a platform that did not directly confront cultural taboos but gently unsettled them ~\cite{murphy2015unsettling, karusala2023unsettling} by showing that conversations around menstruation are valid, answerable, and worthy of care.

Rather than \textit{“making peace”} with constraints, we frame our approach as strategically working within them to create space for gradual change. The chatbot created micro-level openings for young women to question myths and verify information, marking small but meaningful shifts in how they engage with their reproductive health. While such situated interventions cannot claim to overturn patriarchal norms, they show how culturally grounded designs can create fissures in silence and open pathways for incremental change.

Drawing on the synthesized tensions and synergistic principles from Sections 7.1 and 7.2, we articulate a set of design commitments (D1--D4) for LLM-based health chatbots that \emph{design around stigma} rather than outside it in Table~\ref{tab:stigma_framework}. This framework specifies how infrastructural fit, conversational care, expert grounding, and iterative adaptation can be configured to support women’s agency within patriarchal, religiously inflected contexts.

\begin{table*}[t]
\small
\begin{tabular}{p{0.18\textwidth}p{0.26\textwidth}p{0.27\textwidth}p{0.27\textwidth}}
\hline
\textbf{Design Commitment} &
\textbf{Relevance to Designing around Stigma} &
\textbf{Implications for Design} &
\textbf{Trade-offs} \\
\hline

\textbf{D1. Infrastructural fit \& digital \emph{purdah}} &
Embeds the chatbot in infrastructures that already feel religiously and socially permissible, reducing the visibility and social risk of help-seeking in stigmatized domains. &
Deploy on subsidized, widely used platforms (e.g., WhatsApp in Pakistan); provide the option for text interaction; present the system as a ``female'' helper; clearly communicate confidentiality and who can see logs respecting religious norms of \textit{purdah}. &
Platform Dependency undermines design sovereignty, limiting ability to implement custom privacy mechanisms. Framing the system as a “female” helper may also encourage over-reliance on its advice or obscure its non-human, fallible nature.
\\

\hline

\textbf{D2. Conversational care \& localization} &
Uses empathy, validation, and local language practices to make it feel acceptable to ask `unspeakable' questions, while gently questioning harmful norms. &
Adopt local language practices (Roman Urdu mixed with English), recognize cultural (e.g. \textit{`after marriage'} for sexual activity) and religious (e.g. \textit{`exemption from salat'} for menstruation) euphemisms, respond in a non-judgmental tone, acknowledge why myths persist before contrasting them with biomedical explanations, and invite “learning loops” where users can actively test family wisdom against the chatbot’s responses. &
Requires navigating tensions between empathy and medical accuracy (e.g., validating feelings without downplaying risk); overly softened language may obscure clinical urgency, while blunt corrections may feel shaming or dismissive. \\

\hline

\textbf{D3. Expert-grounded, bounded advice} &
Provides an alternative authority to family myths while clearly signalling limits, so that challenging stigma does not create new clinical risks. &
Ground responses in expert-curated content and RAG; state what the chatbot can/cannot do; decline to answer beyond scope (e.g., unknown medicines); redirect to clinicians when needed; cite credible sources (e.g., WHO guidelines) that users can cross-check. &
Balances safety against usability: strict refusals can frustrate users or be perceived as unhelpful, whereas overconfident answers risk clinical harm and may erode trust if contradicted by providers. \\

\hline

\textbf{D4. Iterative, user-in-the-loop adaptation to local logics} &
Treats conversation logs as a lens onto local explanatory models (e.g., hot/cold foods, religious obligations) and updates the system to engage those models rather than dismiss them. &
Analyze logs with clinicians to surface blind spots (e.g., misunderstanding \emph{thanda}/\emph{garam}); add new intents and explanations that bridge between local logics (taseer, prayer) and biomedical models; refine prompts to better handle misspellings and mixed language. &
Trades off learning from real data against privacy and governance concerns, requiring strong consent processes, aggregation, and careful decisions about which user expressions to codify into the system. \\

\hline
\end{tabular}
\caption{A stigma-aware design framework for LLM-based health chatbots in low-resource, patriarchal contexts}\label{tab:stigma_framework}
\end{table*}

\section{Limitations}
Our study has some limitations. The sample was small and demographically narrow, offering rich qualitative insight but limiting generalizability. Our short deployment duration  may not reflect long-term engagement or sustainability of use. Participants’ novelty and curiosity could have shaped their interactions, while issues such as habituation, drop-off in usage, or evolving trust over time remain unexplored. Our study design involved continuous human oversight, including reviewing conversation logs and refining the knowledge base. While this invisible labor was essential for safety and cultural alignment, it limits scalability. The extent to which such oversight can be automated or sustainably maintained in real-world deployments remains an open question.
Finally, we studied the chatbot in a sensitive and taboo-laden domain, where responses can carry heightened social and emotional consequences. Even with expert validation, chatbots cannot replace professional medical advice. Our work should therefore be interpreted as augmenting, not substituting, existing health services.
Together, these limitations underline the need for future research to test culturally sensitive, LLM-based health chatbots across longer timescales, more diverse populations, and broader infrastructural conditions, while continuing to foreground expert validation and community trust.

\section{Conclusion}
\label{section:conclusion}

This paper contributes the first empirical study of a localized LLM-powered chatbot for menstrual health in Pakistan, advancing HCI’s understanding of how conversational systems operate under stigma, silence, and infrastructural constraint. Our study shows that localized AI can create private, judgment-free spaces for exploring sensitive health topics, while also surfacing tensions around trust, cultural alignment, and validation. By showing how language practices, infrastructural limits, and cultural norms shape engagement, we distill design commitments for building safer and more resonant conversational systems. We also highlight the hidden labor of expert validation as essential for cultural relevance and medical safety. This work is not only about correcting hallucinations but about bridging biomedical knowledge with local explanatory models that shape how people understand their bodies. Designing with, rather than against, these models builds trust and makes systems meaningful in everyday life.

Looking ahead, we see opportunities to extend this approach to domains such as fertility, menopause, and sexual health, and to more systematically embed euphemisms and cultural framings into conversational agents. At a broader level, dignity in health technology is not achieved by correct answers alone, but by enabling people to ask questions they could not ask elsewhere, and by receiving responses that resonate with their cultural worlds.

\bibliographystyle{ACM-Reference-Format} 
\bibliography{sample-base}

\appendix
\section{Appendix}

\subsection{Post-deployment Interview Protocol}
\label{protocol}
\begin{enumerate}
    \item Demographic Information
    \begin{enumerate}
        \item What is your name?
        \item What is your age?
        \item What is your educational background?
         \item Do you have a personal phone, a shared phone, or both? 
        \begin{itemize}
            \item If shared, who do you share it with?
        \end{itemize}
    \end{enumerate}

    \item Menstrual Health Information Seeking
    \begin{enumerate}
        \item When was your last visit to the doctor?  
        \begin{itemize}
            \item Have you ever consulted a hakeem or homeopath?
            \item What was your experience with them?
        \end{itemize}
        \item Where do you usually get information about periods?
        \item To what extent do you use digital or technological sources for information about periods?
        \item How do you decide which source to get information from?
        \item How much do you trust each of these sources?  
        \begin{itemize}
            \item How do you validate the information you receive, if at all?
        \end{itemize}
        \item Do you rely on your own experiences for information about your periods?  
        \begin{itemize}
            \item If so, to what extent?
        \end{itemize}
        \item How often do you discuss your periods with friends or family, if at all?
        \item Are there any specific sources you avoid? Why?  
        \begin{itemize}
            \item (Prompt: Ask about any negative experiences in the past)
        \end{itemize}
        \item What usually prompts you to seek information about periods?  
        \begin{itemize}
            \item What sort of queries do you typically focus on (e.g., pain, irregular cycles, diet, medication)?
        \end{itemize}
        \item How do you separate fact from myth regarding periods?  
        \begin{itemize}
            \item For example, have you ever heard any myths (suni sunayi baatain)?
        \end{itemize}
        \item Do you feel that your religion or culture influences your information-seeking behavior?  
        \begin{itemize}
            \item If yes, how?
        \end{itemize}
        \item Are there specific topics (e.g., medication, diet, exercise) you prefer getting information about from specific sources?  
        \begin{itemize}
            \item Please provide examples.
        \end{itemize}
        \item Have you faced any challenges in finding satisfactory (easy to understand, accurate) information about periods?
    \end{enumerate}

    \item Menstrual Health Chatbot
    \begin{enumerate}
        \item What were your initial expectations of the chatbot?
        \item How did you engage with the chatbot (e.g., frequency, types of questions asked)?
        \item How satisfied were you with the responses provided by the chatbot?
        \item Did you trust the responses given by the chatbot?  
        \begin{itemize}
            \item Why or why not?
        \end{itemize}
        \item Did you verify the responses provided by the chatbot?  
        \begin{itemize}
            \item If yes, how?
        \end{itemize}
        \item Did you feel any of the responses were biased or problematic in any way?  
        \begin{itemize}
            \item For example, was there anything said that you didn’t like, or did you have any issues with how the information was presented?
        \end{itemize}
        \item Were there any specific areas where you felt the chatbot performed well?
        \item Were there any areas where you felt the chatbot could improve?
        \item How did your experience with the chatbot compare to other sources of information about periods?
        \item Based on this experience, would you be interested in using chatbots for other types of health education?
        \item How do you manage your privacy on your phone? 
        \begin{itemize}
            \item For example, do you use a password or other methods?
        \end{itemize}
        \item How did you manage your privacy while using the chatbot?
        \begin{itemize}
            \item For example, Did you delete the chats or use any other strategies?
        \end{itemize}
    \end{enumerate}
\end{enumerate}
\subsection{Themes in Menstrual Health FAQs }
\label{theme_FAQs}
\begin{figure*}[ht]
    \centering
    \includegraphics{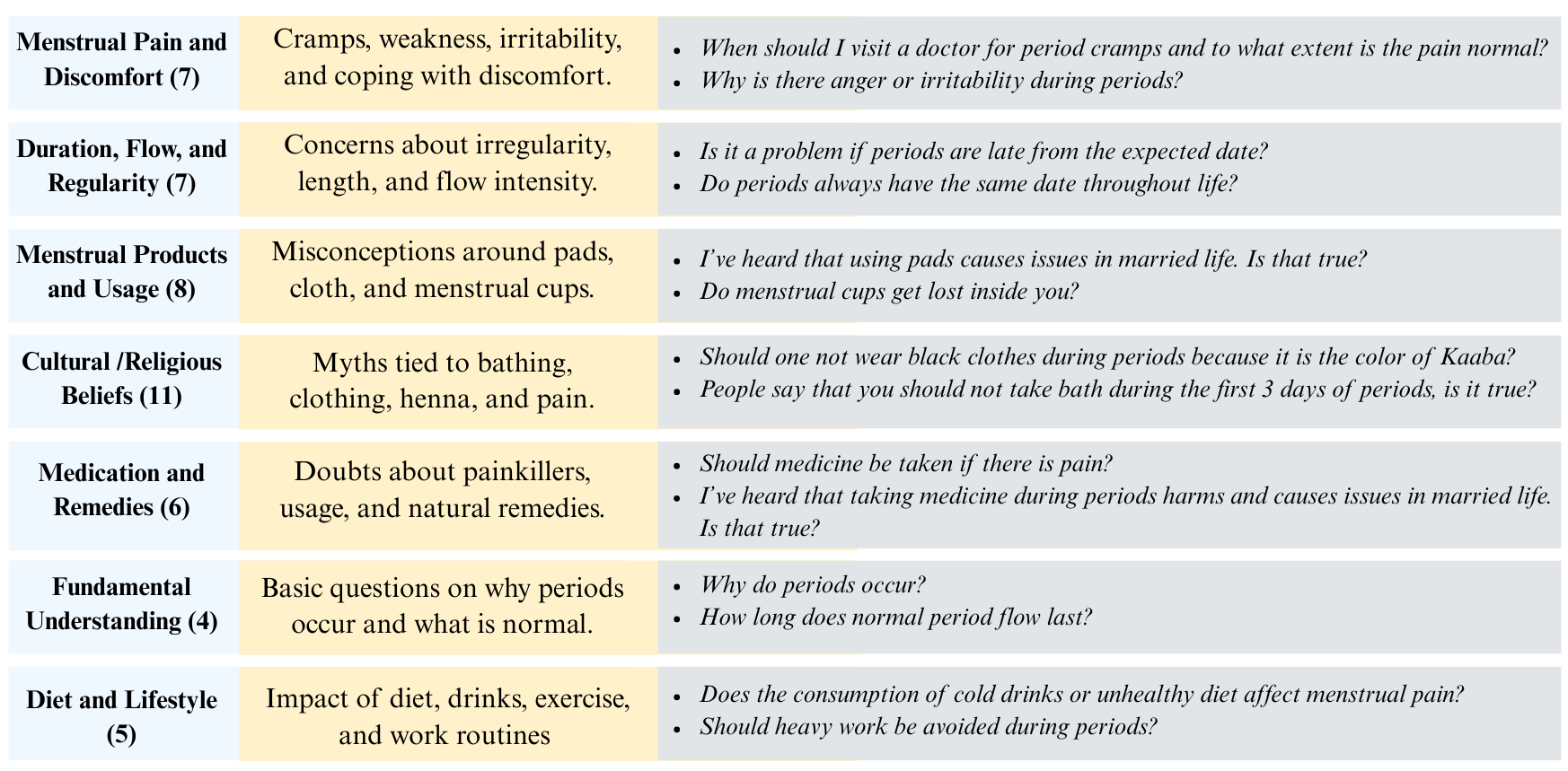}
  \caption{Menstrual Health FAQs gathered from gynecologist and co-design workshops}
    \label{fig:KB}
    \Description{A table of menstrual health FAQs gathered from gynaecologists and co-design workshops. The table has seven categories, each with a short description and example questions. The first category, Menstrual Pain and Discomfort (7), highlights cramps, weakness, irritability, and coping with discomfort, with example questions such as “When should I visit a doctor for period cramps and to what extent is the pain normal?” and “Why is there anger or irritability during periods?” The second, Duration, Flow, and Regularity (7), covers concerns about irregularity, length, and flow intensity, with example questions including “Is it a problem if periods are late from the expected date?” and “Do periods always have the same date throughout life?” The third, Menstrual Products and Usage (8), addresses misconceptions around pads, cloth, and menstrual cups, with examples such as “I’ve heard that using pads causes issues in married life. Is that true?” and “Do menstrual cups get lost inside you?” The fourth, Cultural/Religious Beliefs (11), focuses on myths tied to bathing, clothing, henna, and pain, with questions like “Should one not wear black clothes during periods because it is the color of Kaaba?” and “People say that you should not take bath during the first 3 days of periods, is it true?” The fifth, Medication and Remedies (6), presents doubts about painkillers, usage, and natural remedies, illustrated by questions such as “Should medicine be taken if there is pain?” and “I’ve heard that taking medicine during periods harms and causes issues in married life. Is that true?” The sixth, Fundamental Understanding (4), includes basic questions on why periods occur and what is normal, with examples “Why do periods occur?” and “How long does normal period flow last?” The seventh, Diet and Lifestyle (5), addresses the impact of diet, drinks, exercise, and work routines, with example questions including “Does the consumption of cold drinks or unhealthy diet affect menstrual pain?” and “Should heavy work be avoided during periods?}
\end{figure*}

\subsection{List of Menstrual Health Questions and Myths}
\label{A_FAQs}
\begin{enumerate}
    \item Why is there so much pain during periods?
    \item Why does the body feel lifeless during periods?
    \item Why is there anger or irritability during periods?
    \item Does taking a shower during periods reduce blood flow?
    \item Does wearing black underwear during periods radiate heat?
    \item Which pads should be worn during periods: cotton or Always?
    \item Do menstrual cups get lost inside you?
    \item Are cloth or cotton pads better?
    \item Should medicine be taken if there is pain?
    \item Is excessive bleeding a serious issue during periods?
    \item Is it a concern if periods last more than 7 days?
    \item Is it a problem if periods are late from the expected date?
    \item Does bathing during periods reduce blood flow?
    \item Why do period cramps occur? What should be avoided to prevent cramps?
    \item How long does normal period flow last?
    \item Which pads are good for the environment? I've heard synthetic pads are not good.
    \item Should heavy work be avoided during periods? Is that correct?
    \item Should normal work be continued during periods? Is more exercise beneficial?
    \item Do pores open during periods, affecting circulation?
    \item Why should henna not be applied during periods?
    \item I've heard that taking medicine during periods harms and causes issues in married life. Is that true?
    \item What should be done if there is pain during periods?
    \item What are the benefits and advantages of using pads?
    \item I've heard that using pads causes issues in married life. Is that true?
    \item Which company's pads are good and should be used?
    \item I've heard that using cloth during periods causes rashes. Is that true?
    \item I've heard that if you listen to someone's period pain, it transfers to you. Is that true?
    \item Why do periods occur?
    \item How do people endure pain during periods?
    \item Does using a hot water bottle provide relief during periods?
    \item Should cold water or substances be taken during periods?
    \item When and why do periods occur?
    \item I've heard that you shouldn't bathe during periods because it increases pain. Is that true?
    \item Do periods always have the same date throughout life?
    \item Do certain foods make your first period come sooner?
    \item Does the date of periods change every 3 months?
    \item Should one not wear black clothes during periods because it is the color of Kaaba?
    \item If I don't have menstrual pads, can I use a cloth as a replacement?
    \item When should I visit a doctor for period cramps and to what extent is the pain normal?
    \item Does the consumption of cold drinks or an unhealthy diet affect menstrual pain? Which foods should I avoid during periods?
    \item What are some natural remedies for menstrual pain relief?
    \item Some people say that you should not take a bath during the first three days of periods, is it true?
    \item Can I still do home chores and exercise during my period?
    \item Is it a matter of concern to not have irregular periods?
    \item Is it dangerous to have lighter flow during periods? Does the unclean blood stay inside?
    \item One should keep water contact minimal during periods. Is that true?
    \item Is it advisable to only take medication once throughout the entirety of your menstrual cycle, or can it be taken more often, such as twice or even daily?
\end{enumerate}

\subsection{Example Themes and Codes from Interviews}

\begin{table*}[t]
\small
\centering
\caption{Example themes and codes from the analysis of interviews}\label{tab:sample-codes}
\begin{tabular}{p{3cm}p{4cm}p{8cm}}
\toprule
\textbf{Themes} & \textbf{Codes (Illustrative subset)} & \textbf{Example participant quotes} \\
\midrule
Navigating layered information sources &
Mother as primary trusted source &
\emph{“I mainly ask my mom… I will only trust the information if she says it’s correct.”} (C13)\\[0.6em]

 & Cross-checking online information with family or doctors &
 
\emph{“Whatever YouTube tells you, you have to confirm it first. You never know what’s being told is actually true. I always cross-check whatever they say in a YouTube video from different sources.”} (C11)\\[0.6em]

Reworking Menstrual Norms, Religion, and Secrecy &
Navigating myths through search and experience  &
\emph{“Either I go to my mother or I directly go to Dr. Bilquees [YouTube Channel]...I have never had a bad experience with Dr. Bilquees so I never asked anyone else.”} (C1)\\[0.6em]

 & Uncertainty around religious practices & \emph{“They say that you should not touch a prayer mat when you’re on your period. But I don’t know about this, whether it’s actually true or not.”} (C3)\\[0.6em]

 & Secrecy around menstruation &
\emph{“Mom used to ask me not to tell anyone outside, just tell me. But as I grew older, I got more information about periods, and the embarrassment kept decreasing. ”} (C9)\\[0.6em]

Chatbot as confidential, responsive companion &
Immediate responses &
\emph{“ I thought that I’d get my reply after a while. But I kept getting replies immediately. I liked it. Reply was given quickly. ”} (C1) \\[0.6em]

 & No hesitation using chatbot &
\emph{““When I asked it [chatbot], I didn’t have any hesitation. I asked all the questions easily and I got the
answers easily. That’s why I didn’t have to ask anyone else. You can’t ask your family and doctor about
certain things.”} (C8)\\[0.6em]

\bottomrule
\end{tabular}
\end{table*}

\subsection{Pilot Deployment and System Refinements}
\label{app:pilot_refinements}
The pilot phase served as an initial field test to surface challenges in both technical functionality and the accuracy of LLM-generated responses. During one week pilot with seven participants (130 messages), no hallucinations or harmful responses were identified. However, users noted response delays and the lack of confirmation cues, which left them uncertain whether their messages were received or processed.
Based on user feedback, we reconfigured our backend server to reduce the delays, added feedback mechanism, and refined the LLM module as follows:
\begin{itemize}
    \item {\textit{\textbf{Upgraded Health Assistant.}}} After the pilot, we transitioned to the OpenAI Assistant API V2, which offered finer control over the RAG. We set chunk sizes so each one matched a single gynecologist-validated question-answers, which kept each question-answer pair separate and made retrieval cleaner. We then varied the number of retrieved chunks (top-k) to balance answer accuracy against latency, selecting a small k that reliably surfaced the right answer without overloading the prompt. Using both embedding-based similarity and keyword matching improved recall on short, colloquial queries (e.g., misspellings, Roman-Urdu variants) while still catching exact medical terms; this combination increased the chance that the most relevant question-answer chunk appeared in the top results.
    In addition, we selected the GPT-4o model, which provided stronger language comprehension and faster responses. We refined the assistant’s instructions to emphasize menstrual and female health, cultural sensitivity, and concise delivery.

\item {\textit{\textbf{Enhanced Knowledge Base.}}} We rephrased each question in the knowledge base multiple times, both in English and Roman Urdu This is particularly useful in the context of RAG because it increases the likelihood that the retriever model can
match the user’s query with the most relevant document chunks regardless of how the query is formulated.

\item {\textit{\textbf{Read Receipts.}}} We added WhatsApp read receipts to reassure users that their queries had been received and were being processed, aiming to reduce confusion.

\item {\textit{\textbf{Prompt Inclusion}}} Along with the earlier system instructions, we appended a well-defined prompt to each user query to guide the LLM in generating responses. This prompt acted as a condensed version of the assistant’s instructions, reinforcing key guidelines on scope, cultural sensitivity, and brevity. Prior work shows that repeating instructions can help LLMs stay focused on critical aspects of a task. By reiterating this guidance within the prompt, we aimed to ensure that the model’s outputs were more aligned with our design goals.
\end{itemize}

\subsection{Linguistic Variations Used in Participant Interactions}
    \label{var}

\begin{figure*}[ht]
    \centering
    \includegraphics[width=0.75\textwidth]{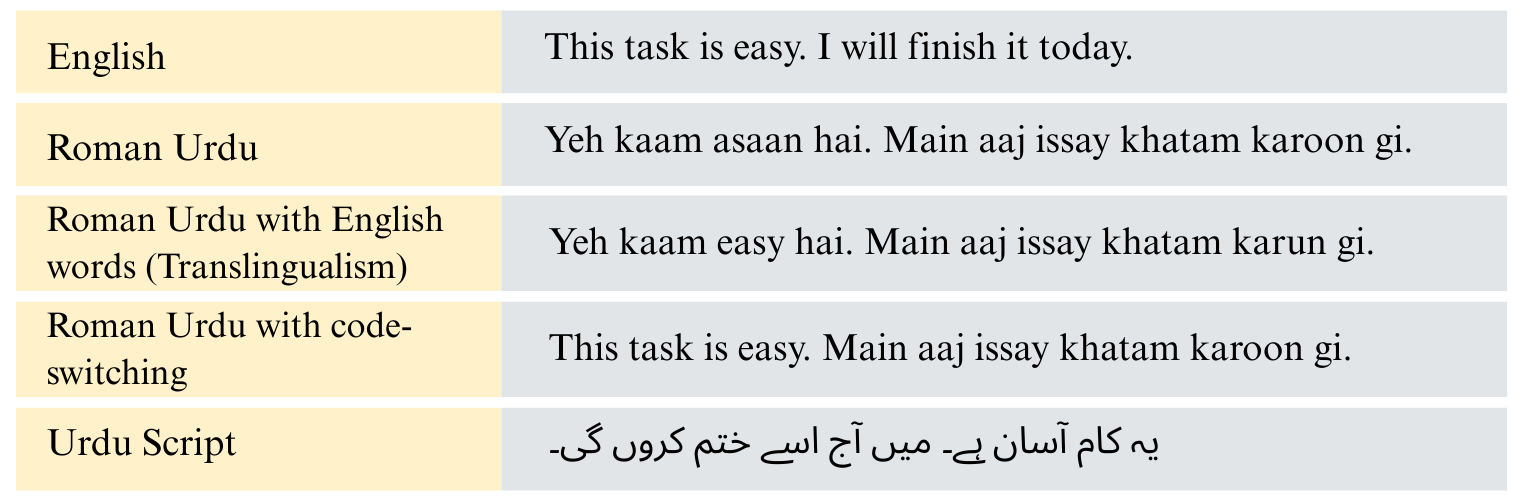}
  \caption{An example illustrating multiple variations of the same sentence}
    \label{fig:var}
    \Description{A table with five rows, each presenting the same sentence in different language variations. The English row reads: This task is easy. I will finish it today. The Roman Urdu row reads: Yeh kaam asaan hai. Main aaj issay khatam karoon gi. The Roman Urdu with English words row reads: Yeh kaam easy hai. Main aaj issay khatam karun gi. The Roman Urdu with code-switching row reads: This task is easy. Main aaj issay khatam karoon gi. The Urdu Script row shows the same sentence in Urdu script.}
\end{figure*}

\subsection{Sample Screenshots of Chatbot Coversations}
\label{sample_convo}

\begin{figure*}[h!t]
  \centering
  \includegraphics[width=0.6\textwidth]{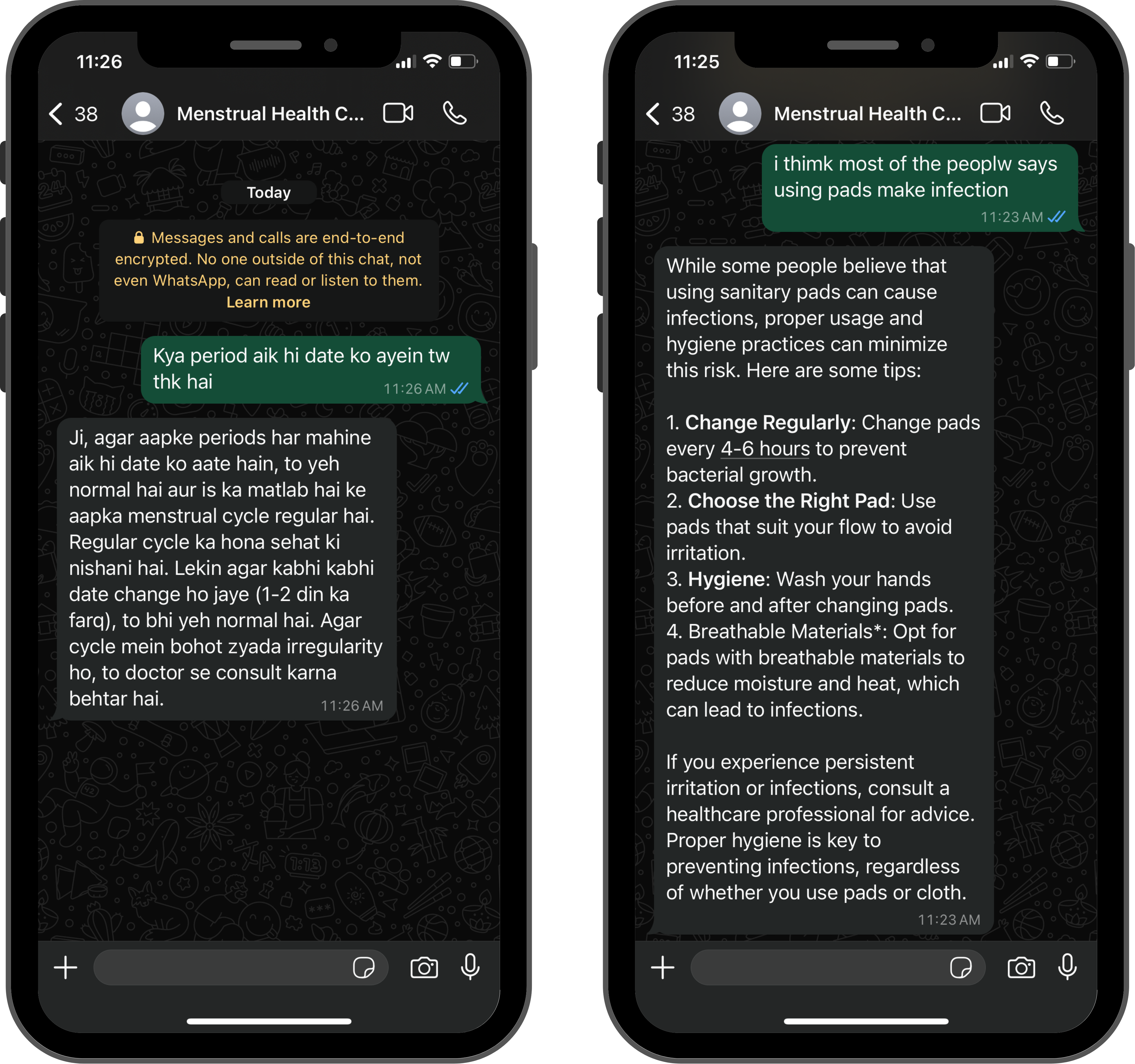}
  \caption{Left: Participant asking a question from Chatbot in Roman Urdu.
Participant: "Is it okay if my period always starts on the same date each month?"
Chatbot: "Yes, if your periods always start on the same date each month, this is normal and indicates that your menstrual cycle is regular. A regular cycle is a sign of good health. However, if the date occasionally changes (1-2 days difference), it is still normal. If there is significant irregularity in the cycle, it is better to consult a doctor."
  \\Right: Participant discussing concerns about using pads with the Chatbot in English.}
    \Description{Screenshot showing a WhatsApp conversation. On the left, the participant asks a question in Roman Urdu about whether it is normal for her period to start on the same date each month. The chatbot responds in Roman Urdu, explaining that it is normal and a sign of a regular menstrual cycle. On the right, the participant raises a concern in English about using pads causing infections. The chatbot replies in English, acknowledging that some people believe pads can cause infections. It provides advice on minimizing the risk by following proper hygiene practices, such as changing pads regularly, using the right pads, maintaining hygiene, and opting for breathable materials. The chatbot suggests consulting a healthcare professional if infections persist.}
    \label{convo_combined}

\end{figure*}

\end{document}